\documentclass[twocolumn,showpacs,superscriptaddress,preprintnumbers,amsmath,amssymb,prc]{revtex4-1}
\usepackage{float}
\usepackage{placeins}
\usepackage{graphicx}
\usepackage{dcolumn}
\usepackage{bm}
\usepackage{ifpdf}
\usepackage{epstopdf}
\usepackage{slashed}
\usepackage{amsfonts}
\usepackage{mathrsfs}
\usepackage{amssymb}
\usepackage{multirow}
\usepackage{CJK}
\usepackage{color}
\usepackage{subfigure,dcolumn}
\usepackage[T2A,T1]{fontenc}
\usepackage[russian,english]{babel}
\usepackage{amsmath}
\usepackage{listings}
\usepackage{booktabs}
\usepackage{youngtab}
\usepackage{hyperref}

\begin{document}

\title{Machine learning the impact parameter in heavy-ion collisions at $\sqrt{s_{\rm NN}}$ = 4 and 11 GeV: a cross-check study with UrQMD, AMPT, and JAM}

\author{Xiaoqing Yue}
\affiliation{Institute of Theoretical Physics, Shanxi University, Taiyuan 030006, China}
\affiliation{School of Science, Huzhou Normal University, Huzhou 313000, China}
\author{Guojun Wei}
\affiliation{Institute of Theoretical Physics, Shanxi University, Taiyuan 030006, China}
\affiliation{School of Science, Huzhou Normal University, Huzhou 313000, China}
\author{Yongjia Wang}
\affiliation{School of Science, Huzhou Normal University, Huzhou 313000, China}

\author{Zhilong Li}
\affiliation{China Institute of Atomic Energy, Beijing 102413, China}
\affiliation{School of Science, Huzhou Normal University, Huzhou 313000, China}
\author{Pengcheng Li}
\affiliation{School of Science, Huzhou Normal University, Huzhou 313000, China}
\author{Haojie Xu}
\affiliation{School of Science, Huzhou Normal University, Huzhou 313000, China}
\author{Xiangrong Zhu}
\affiliation{School of Science, Huzhou Normal University, Huzhou 313000, China}
\author{Qingfeng Li}
\email[Corresponding author, ]{liqf@zjhu.edu.cn}
\affiliation{School of Science, Huzhou Normal University, Huzhou 313000, China}
\author{Fuhu Liu}
\affiliation{Institute of Theoretical Physics, Shanxi University, Taiyuan 030006, China}

\author{Yasushi Nara}
\affiliation{Akita International University, Yuwa, Akita-city 010-1292, Japan}

\date{\today}

\begin{abstract}

By generating heavy-ion collision data with the ultrarelativistic quantum molecular dynamics (UrQMD) model, a multiphase transport (AMPT) model, and the JAM model, the impact parameter ($b$) in Au+Au collisions at $\sqrt{s_{\rm NN}}$ = 4 and 11 GeV is reconstructed using supervised learning and unsupervised learning in machine learning (ML). In supervised learning, the performance of ML algorithm is cross-checked by using data obtained from these three transport models. It is found that the typical mean absolute error (MAE) which measures the average magnitude of the absolute difference between the true and predicted $b$ is between 0.2-0.4 fm, even when training ML algorithm with data generated from one model but testing with data from others. While the conventional method (i.e., a polynomial fit to multiplicity as a function of $b$) only works for data generated from the same model. In the classification task, the present ML-based method also shows significantly superior results compared to the traditional approach. In unsupervised learning, the K-means clustering algorithm is used to partition collision events directly from experimental-style observables, showing that the algorithm autonomously identifies six clusters corresponding to different centrality classes without relying on predefined model-based binning. Our study demonstrates the strong robustness of using an ML algorithm trained on transport-model data for impact-parameter determination, and indicates that this method has the potential to be generalized to handle real experimental data.

\end{abstract}

\maketitle
\section{Introduction}
One of the primary scientific goals of heavy-ion collisions (HICs) at the center-of-mass energy of a few GeV is to explore the properties of dense nuclear matter. This is crucial to understanding the phase diagram of strongly interacting matter, the dynamics of neutron star mergers and HICs, the structure and properties of neutron stars \cite{Danielewicz:2002pu, Fukushima:2010bq, Sorensen:2023zkk}. To this end, several experimental programs have been performed or are planned, such as the CSR external-target experiment (CEE) at the Heavy Ion Research Facility at Lanzhou (HIRFL) \cite{Guo:2024zij}, the Compressed Baryonic Matter experiment (CBM) at the Facility for Antiproton and Ion Research (FAIR) \cite{Klochkov:2021eyo}, the STAR Fixed-Target Program at Relativistic Heavy Ion Collider (RHIC) \cite{STAR:2017sal}, and the Multi-Purpose Detector (MPD) of the collider experiment at Nuclotron-based Ion Collider fAcility (NICA) \cite{MPD:2022qhn, Syresin:2025lye}. The properties of dense matter cannot be measured directly in experiments but are deduced from the comparison of observables simulated with transport models to that with experimental measurements. Usually used observables include the yields and energy spectra of charged particles, the collective flow \cite{Zhu:2024tns}, the mean transverse kinetic energy or transverse momentum, as well as various correlations and fluctuations \cite{TMEP:2022xjg, STAR:2013gus, Zhang:2025ale}. All of these observables are strongly related to the collision centrality or the impact parameter $b$ which is defined as the perpendicular distance between the centers of the two colliding nuclei in the classical picture, thus determination of impact parameter in heavy-ion collision experiments is of great importance. However, $b$ is not directly measurable in experiments, as its typical value is only a few femtometers. 
Centrality or $b$ of a collision can only be inferred from the distribution of an experimental observable \cite{Cavata:1990gk, Miller:2007ri, HADES:2017def, Zhang:2025ale}. 

In relativistic heavy ion collisions, the total charged multiplicity has been widely used to determine the centrality in experiments, with the help of the Glauber model to fit measured multiplicity distributions \cite{STAR:2009sxc, STAR:2012och, STAR:2019vcp, STAR:2019bjj, STAR:2022etb}. The Glauber model assumes that all nucleons travel in a straight line along the beam axis based on the eikonal approximation, and it neglects the transverse motion of nucleons during the short passage time of the two high-energy nuclei. As discussed in Ref. \cite{HADES:2017def}, the Glauber model at higher energies (i.e., $\sqrt{s_{\rm NN}}$ around 10 GeV and above) is well validated because of the satisfaction of the eikonal approximation, but modifications are needed when applying it in HICs at lower energies.
Our recent study shows that different centrality determination methods can lead to significantly different impact-parameter distributions of selected events, which in turn induce non-negligible variations in final-state observables. In particular, the Glauber model may become unreliable at intermediate energies, where the associated uncertainties can be comparable to or even exceed those originating from the nuclear equation of state (EoS) \cite{Yue:2026flb}.
In HICs at low and intermediate energies (beam energy of a few hundred MeV per nucleon), the determination of centrality is more challenging and the situation is more complex than in relativistic HICs. Because the de Broglie wavelength is comparable with the size of the nucleus, the classical picture is no longer valid. Consequently, most experimental studies have only focused on very central or semi-central collisions, i.e., the considered range of reduced impact parameter ($b_0=b/b_{max}$) does not significantly exceed $b_0=0.5$. Here $b_{max}$ is the sum of hard-sphere radii of the target and projectile nuclei in the geometrical prescription of Ref. \cite{Cavata:1990gk}. 
Usually, $b_{max}$ can be calculated via the relation $b_{max} = r_0 \times (A_T^{1/3}+A_P^{1/3})$ with $r_0$ = 1.12-1.20 fm. $A_T$ and $A_P$ are the mass numbers of the target and projectile nuclei, respectively. In some experimental studies, $b_{max}$ is not directly calculated but is deduced from experimental measurements \cite{Lukasik:1997zz}. Several observables have been proposed to infer centrality in HICs at low and intermediate energies, including the total transverse kinetic energy of light charged particles, the total charge bound in fragments, the ratio of transverse-to-longitudinal kinetic energy, and so on \cite{Lukasik:2004df, Schuttauf:1996ci, FOPI:1996cjz, LeFevre:2015paj}.

Nowadays, HICs at the center-of-mass energy of a few GeV, i.e., the energies covered by STAR-FXT, MPD at NICA, and CBM at FAIR, have attracted a lot of theoretical and experimental attention. The determination of centrality at this energy regime is of great significance for exploring the location of a possible critical end point. 
In this work, $\sqrt{s_{NN}}$=4 and 11 GeV are considered as two representative benchmark energies in this regime: the former lies deeper in the high-baryon-density, predominantly hadronic domain, while the latter is located near the upper edge of the low-energy BES/NICA window and provides a complementary reference point with different particle-production dynamics and model sensitivities \cite{Luo:2017faz, Lbchen, Werner:2024ntd}. 
Simulations at this energy regime are highly model-dependent, unlike the case at very high energies (i.e., LHC and RHIC energies), where most of experimental observables can be well reproduced by theoretical models. Thus using the Glauber model to establish the relationship between charged multiplicity and impact parameter will be model dependent and may not be valid. In recent years, the application of machine learning (ML) in scientific research has received unprecedented attention, and prodigious progress has been made. Due to the strong capability of ML algorithms to process and analyze data, and to decode information from data, ML has emerged as a novel tool for extracting information in many areas of science. 
For example in nuclear physics, ML has been successfully employed to refine nuclear mass models \cite{qu2025nuclear,Mumpower:2022peg,gao2021machine,Yuksel:2024zky}, to decode information on the nuclear EoS \cite{Kvasiuk:2020izb,Wang:2023kcg,Patra:2025xtd,Wang:2024ahj}, to optimize the parameters of the Woods-Saxon potential \cite{Pu:2023jae,Gao:2023jjf,Wu:2025txu}, to predict nuclear charge radius \cite{dong2023machine,li2025machine,Guo:2026bfz,Li:2024tth}, low-lying nuclear excitation properties \cite{li2026study}, nuclear binding energy \cite{Du:2023hcq,Huang:2025ubz,Bentley:2024bnm,Yuan:2024ivv}, nuclear half-life properties \cite{Jyothish:2025ddz,Cai:2023mak,Ma:2023ofi,Tang:2024xxj,Shree:2025hyu,Niu:2018trk}, and nuclear reaction cross-sections \cite{Li:2023ukd,Choi:2024nkr}, etc. 
To determine the impact parameter, the use of machine-learning techniques can be traced back to the pioneering studies in the 1990s. In Refs.~\cite{Bass:1993vx, Bass:1996ez}, Bass \textit{et al.} developed neural-network-based methods for event-by-event impact-parameter estimation in intermediate-energy HICs, using input data generated from IQMD model simulations, thereby demonstrating improved accuracy over conventional methods. Since then, various machine-learning algorithms, including support vector machines (SVMs) \cite{DeSanctis:2009zzb}, neural networks (NNs) \cite{David:1994qc,Galaktionov:2023dui,Galaktionov:2023ilx, Zhang:2021zxd}, multilayer perceptrons (MLPs) \cite{Xiang:2021ssj}, boosted decision trees \cite{Mallick:2021wop}, and Bayesian- or clustering-based methods \cite{LiLI:2022vlp}, have been employed for impact-parameter reconstruction and centrality determination in heavy-ion collisions.

In our previous works, three commonly used ML algorithms, i.e., the artificial neural network (ANN), the convolutional neural network (CNN), and the light gradient boosting machine (LightGBM), were used to determine the impact parameter by analyzing data generated from the ultrarelativistic quantum molecular dynamics (UrQMD) model, and good performance was achieved in HICs at beam energies from 0.2 to 1 GeV$/$nucleon \cite{Li:2021plq,Li:2020qqn}. The aim of the present work is to study the capability of ML algorithms in the determination of 
impact parameter in HICs at STAR-FXT and NICA  energies and to discuss the possible effect of model-dependent data. To this end, three transport models, i.e., the UrQMD model, the AMPT model, and the JAM model are used to generate labeled data.

The rest of this paper is organized as follows. In Sec. II, we
introduce the methodology that we use in the present work,
including the labeled data generated with transport models, and the ML algorithms. The results are discussed in detail in Sec. III. The conclusions
are given in Sec. IV.

\section{Methodology}
To infer the impact parameter or centrality in heavy-ion collisions with ML algorithms, one needs data and ML algorithms. Experiments of heavy-ion collisions and simulations with transport models are two ways to generate data. Since the data generated from transport model simulations is relatively easier to control and can be used to evaluate the performance of ML algorithms,  the present work adopts data generated by the transport models. In this section, we describe how the data are generated and the ML algorithms used in the present work.

\subsection{Data generation}
\begin{figure}
    \centering
    \includegraphics[width=\linewidth]{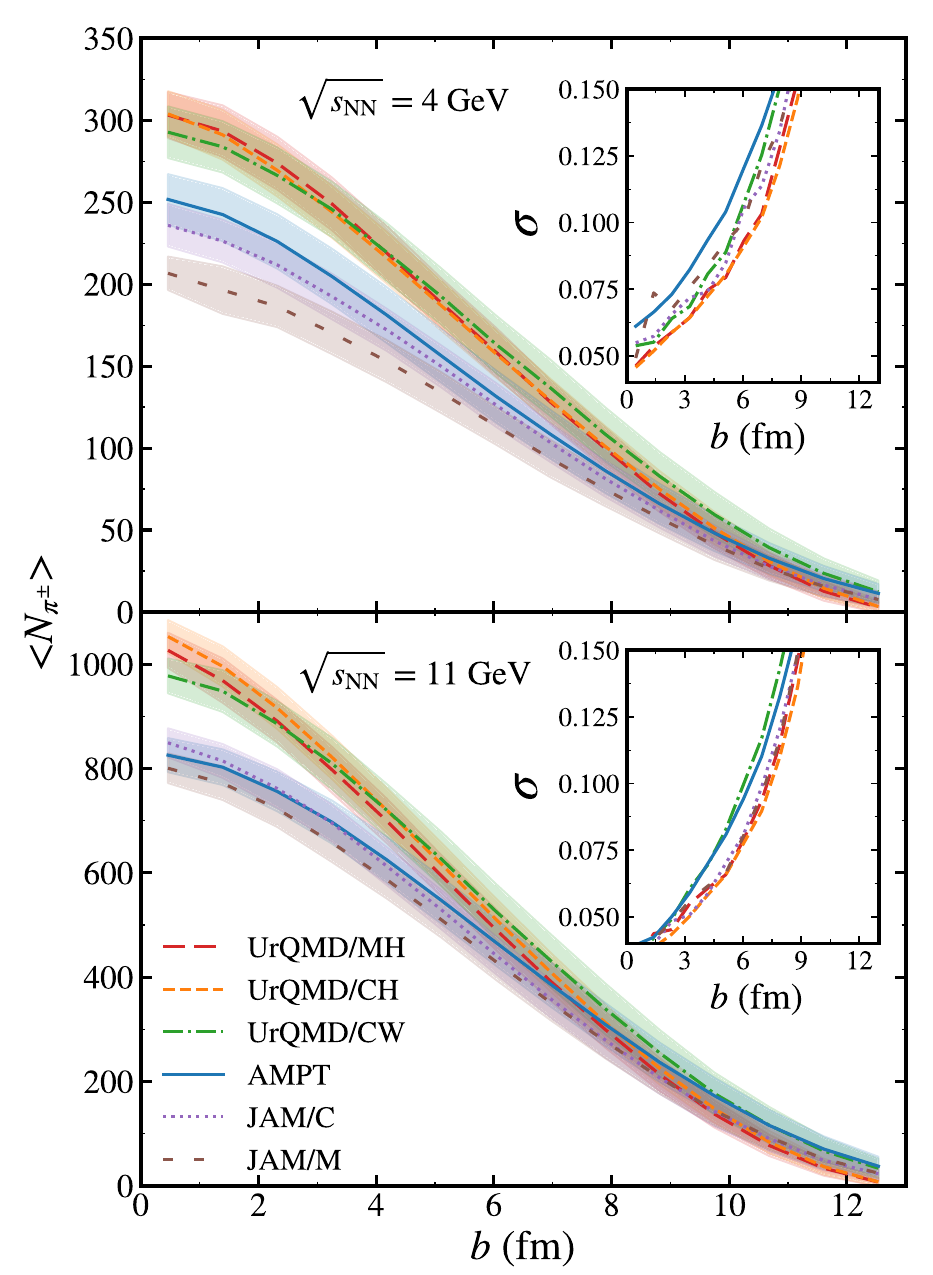}
    \caption{The average number of charged $\pi$ ($\langle N_{\pi^{\pm}}\rangle$) produced in Au + Au collisions at $\sqrt{s_{\rm NN}}$=4 (upper panel) and 11 GeV (lower panel) is plotted as a function of impact parameter. The curves and shaded bands denote the mean value and the standard deviation, respectively. The standard deviation denotes variation from event to event. The inset plot shows the impact parameter dependence of the ratio between the standard deviation and $\langle N_{\pi^{\pm}}\rangle$.}
    \label{fig:1}
\end{figure}

\begin{figure}
    \centering
    \includegraphics[width=\linewidth]{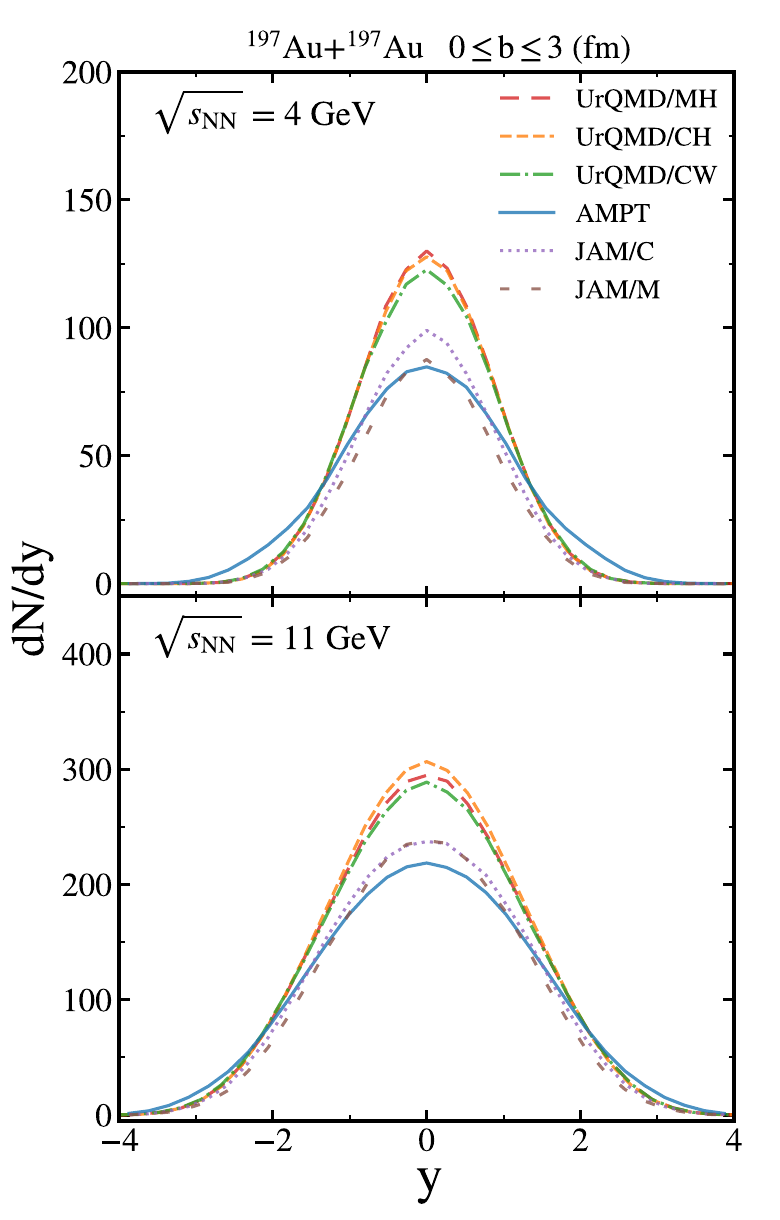}
    \caption{The rapidity distribution of charged $\pi$ produced in central Au + Au collisions at $\sqrt{s_{\rm NN}}$=4 (upper panel) and 11 GeV (lower panel). }
    \label{fig:2}
\end{figure}
The UrQMD model, AMPT model, and JAM model are used to generate labeled data \cite{Li:2011zzp, Wang:2021symmetry, Gao:2023plb, Lin:2004en, Nara:1999dz, Isse:2005nk, Nara:2021fuu}. It is known that for studying HICs at several GeV, the three main ingredients (i.e., the initialization \cite{Yue:2022zfu, Li:2011zzp}, the mean-field propagation \cite{Xu:2013sta, Nara:2016hbg, Nara:2021fuu, Li:2018wpv} and collision terms \cite{Nan:2024ogc}) of transport model all may play a role in the simulation of observables. To generate data from models with the largest parameter space, the UrQMD model with different sets on the above three ingredients is employed. In the initialization of UrQMD model, two (i.e., hard-sphere and Woods-Saxon) different modes are used. In the mean-field propagation term, we consider two modes, i.e., pure cascade mode and mean-field mode. Together, these form three setups: UrQMD/MH (mean-field and hard-sphere), UrQMD/CH (cascade and hard-sphere), and UrQMD/CW (cascade and Woods-Saxon). In addition, the string-melting version of AMPT model, as well as the JAM model with Woods-Saxon initialization in cascade mode (JAM/C) and in mean-field mode with a soft EoS (JAM/M), are also used. Usually, observables calculated with different models or parameter choices are very different from each other. As shown in Figs. \ref{fig:1} and \ref{fig:2}, both the yield of charged $\pi$ and its rapidity distribution obtained from different models exhibit considerable divergence. Especially for the yield of charged $\pi$ at mid-rapidity, the result obtained with UrQMD/CH is about 60\% larger than that of AMPT at $\sqrt{s_{\rm NN}}$ = 4 GeV. Fig. \ref{fig:1} also illustrates the standard deviation and the ratio between the standard deviation and the mean value of charged $\pi$ yield. These two quantities partly estimate how strong the fluctuation between different events (with the same impact parameter) is. As can be seen in the inset plot of Fig. \ref{fig:1}, the fluctuations of the yield of charged $\pi$ in UrQMD/MH and UrQMD/CH are relatively weaker than those in UrQMD/CW and AMPT. Apparently, stronger fluctuations make it more difficult to determine the impact parameter. 
We note here that previous studies based on the UrQMD model have shown that, after suitable improvements of the model ingredients, pion-related observables such as the charged-pion multiplicity and the ($\pi^-/\pi^+$) ratio can reasonably describe the main experimental trends at low and intermediate energies \cite{Liu:2025pzr, Liu:2020jbg}. Our purpose is not to describe the experimental data but instead to generate labeled data from a larger model space and to evaluate the performance of ML algorithms on different data, thus the above mentioned six modes are used to generate data. 

For each mode, ten thousand Au + Au collision events with known impact parameter 0$<b<$13 fm at $\sqrt{s_{\rm NN}}$ = 4 and 11 GeV are simulated. The data of 6000 events are set as training data, and the remaining 4000  events are set as test data. For each event, six features, the yield of charged pions ($N_{\pi^-}$ and $N_{\pi^+}$), the yield of charged pions at mid-rapidity ($n_{\pi^-}$ and $n_{\pi^+}$), the total transverse momentum of charged pions ($p_{t {\pi^-}}$ and $p_{t {\pi^+}}$), are calculated. These quantities are known to be closely related to the impact parameter, but there is no one-to-one correspondence between them because the process of HICs involves large fluctuations from event to event.
The above quantities are selected not only because they are measurable in HIC experiments on the event-by-event basis but also because they have relatively smaller event-by-event fluctuations. In principle, observables like the large fragment with projectile (target) rapidity and charged-particle multiplicity are also good candidates for determining the impact parameter, but transport model simulations of these observables are highly model-dependent, so they are not used in the present work.

\subsection{Machine learning algorithms }

LightGBM is an advanced ML framework developed by Microsoft. It is widely used in ML competitions and industrial applications due to its superior performance compared to other algorithms. LightGBM employs several optimization techniques to significantly reduce training time while maintaining high accuracy. Gradient-based one-side sampling (GOSS),  exclusive feature bundling (EFB), and leaf-wise growth are main optimization techniques. GOSS keeps instances with large gradients (misclassified/hard-to-learn samples) and randomly drops some with small gradients (well-learned samples), which reduces computation time without significantly sacrificing accuracy. EFB is designed to reduce feature dimensionality, thereby improving training speed and memory efficiency, and is particularly suited for high-dimensional sparse data. Leaf-wise growth prioritizes expanding the leaf node that currently delivers the greatest loss reduction, rather than mechanically splitting all nodes layer by layer (level-wise growth approach). The advantages of LightGBM include: 
(1) faster training efficiency, (2) lower memory usage, (3) higher accuracy, and (4) ability to tackle large-scale data. In our previous works, LightGBM has shown great promise for handling several issues in nuclear physics, such as prediction of nuclear mass and excitation energies of low-lying states \cite{gao2021machine}, the determination of impact parameter \cite{Li:2021plq,Li:2020qqn}, and constraining the stiffness of the nuclear symmetry energy \cite{Wang:2022cda}. In addition, our previous work has shown that for feature data (represented by numerical or categorical features), LightGBM generally outperforms other algorithms. Thus, in the present work, LightGBM is used, and its hyperparameters, such as num\_boost\_round (maximum
number of decision trees allowed), num\_leaves
(maximum number of leaves allowed per tree), max\_depth (maximum depth allowed per tree) are set to their default values. We
checked that varying these parameters did not significantly alter the results.
 
\section{Results}\label{sec:artwork}

Impact parameter identification can be set as supervised learning and unsupervised learning tasks. In supervised learning, data with the corresponding input and output pairs are required, allowing research on regression and classification tasks. In the regression task, the output is the value of impact parameter for each event, while in the classification task, the goal is to determine which centrality class each event belongs to, so the output is a discrete class label. In the unsupervised learning task, all events can be partitioned into several predefined categories to achieve the purpose of centrality classification. In this section, results of both supervised learning and unsupervised learning tasks are discussed. 

\subsection{Regression task}

\begin{figure}
    \centering
    \includegraphics[width=0.9\linewidth]{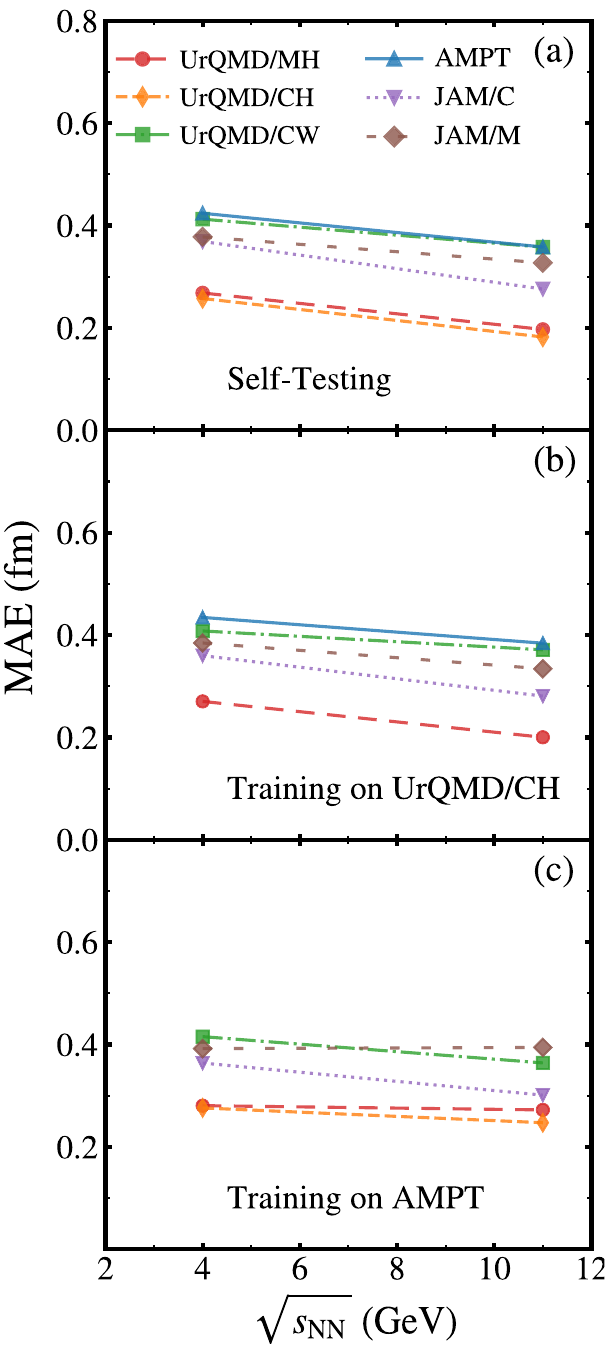}
    \caption{The MAE is plotted as a function of $\sqrt{s_{\rm NN}}$. Panel (a) shows the results when the training and the test data are generated from the same model. Panel (b) and (c) are the results when the training data are generated from UrQMD/CH and AMPT, respectively. }
    \label{fig:3}
\end{figure}

\begin{figure*}
    \centering
    \includegraphics[width=\linewidth]{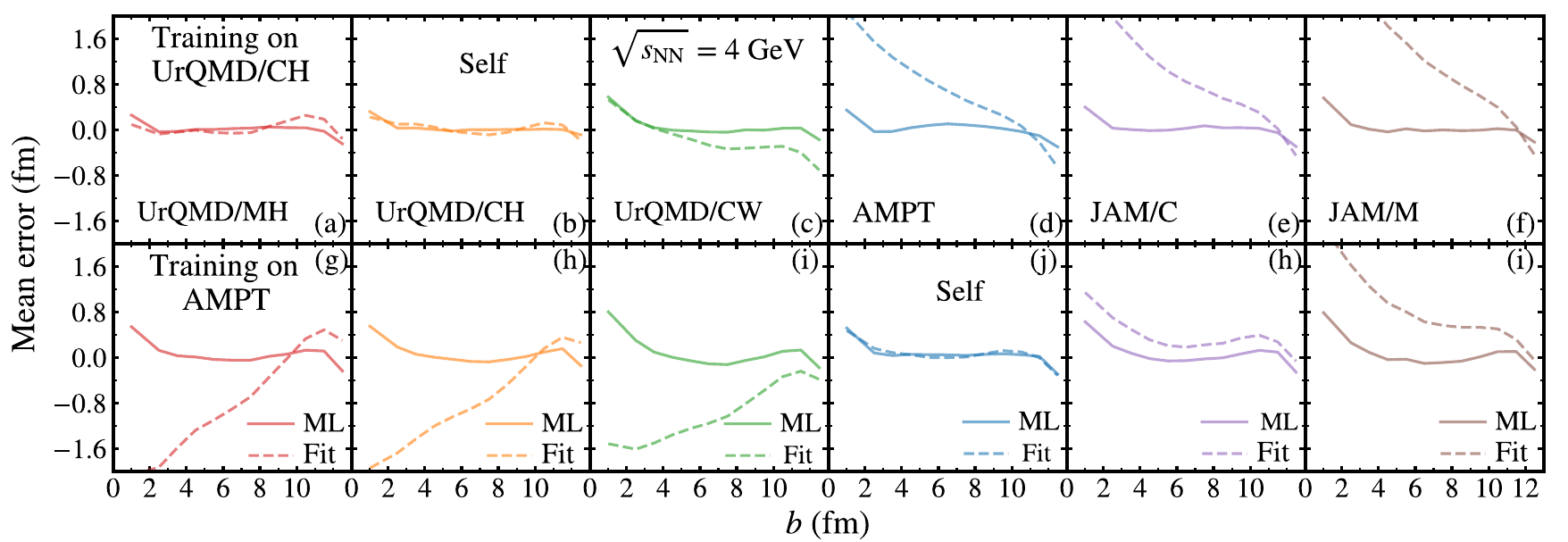}
    \caption{Mean error of estimated impact parameter as a function of impact parameter at $\sqrt{s_{NN}}$ = 4 GeV. Dashed line denotes the result obtained with polynomial fitting method, the solid line denotes the result obtained with ML approach. In the upper panel, both the ML algorithms and polynomial fitting are trained with data generated from UrQMD/CH; in the bottom panel, the training data are generated from AMPT.   
    }
    \label{fig:4}
\end{figure*}

\begin{figure*}
    \centering
    \includegraphics[width=\linewidth]{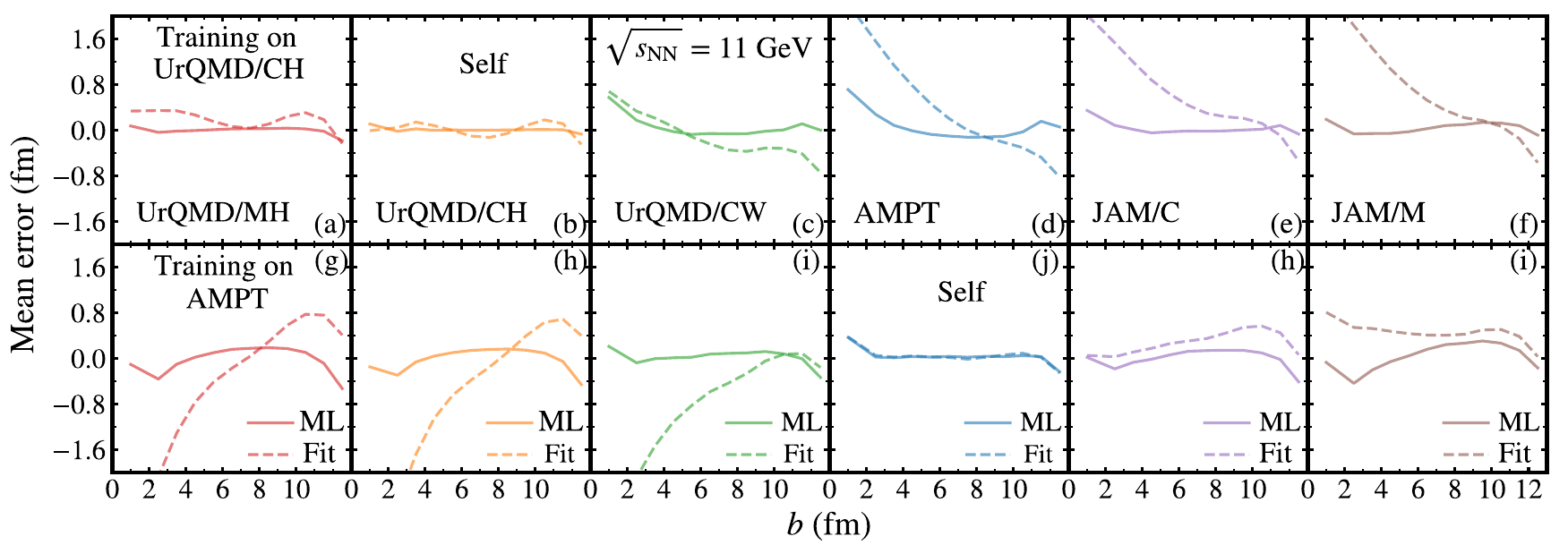}
    \caption{Same as Fig. \ref{fig:4}, but for the mean error of estimated impact parameter as a function of impact parameter in $\sqrt{s_{NN}}$ = 11 GeV.}
    \label{fig:5}
\end{figure*}

As for data generated with transport model, the value of impact parameter for each event is known, one can use the mean absolute error (MAE) to evaluate the performance of ML algorithms, which reads,
\begin{equation}\label{eq2}
\text { MAE } = \frac{1}{N} \sum_{i=1}^{N} \left| b_i ^{\text{pred}}-b_i ^{\text{true}}\right|
\end{equation}
Here $N$ is the number of tested events, $b_i ^{\text{pred}}$ and $b_i ^{\text{true}}$ are the predicted and true impact parameter of each event, respectively. 

The results on MAE for different scenarios are displayed in Fig. \ref{fig:3}. It can be seen that the typical values of MAE are around 0.2-0.4 fm even for scenarios when training and test data are generated from different models, demonstrating the strong ability of ML algorithms in determining impact parameter. In the scenario of self-testing, the MAE values of UrQMD/MH and UrQMD/CH are the smallest, the results of UrQMD/CW and AMPT are the largest, and that of JAM/C and JAM/M lie between these two groups. This ordering corresponds to the trend of the ratio $\sigma$ observed in Fig. \ref{fig:1}. The initialization in UrQMD/MH and UrQMD/CH is identical, while that in UrQMD/CW and AMPT (both using the Woods-Saxon scenario) is quite similar. It can be seen in Fig. \ref{fig:3} that the MAE values of UrQMD/MH and UrQMD/CH are very close, and so are those of UrQMD/CW and AMPT. In addition, JAM/C and JAM/M both adopt Woods-Saxon initialization, and their MAE values are also relatively close, especially at 4 GeV. These results show the importance of initialization when studying the impact parameter with ML. It is known that, with different treatments of the initialization, both the initial density distribution and fluctuations may differ significantly, thus resulting in different magnitudes of fluctuations on final observables. Apparently, stronger fluctuations in observables tend to produce a larger MAE.

\begin{figure*}[!t]
    \centering
    \includegraphics[width=\linewidth]{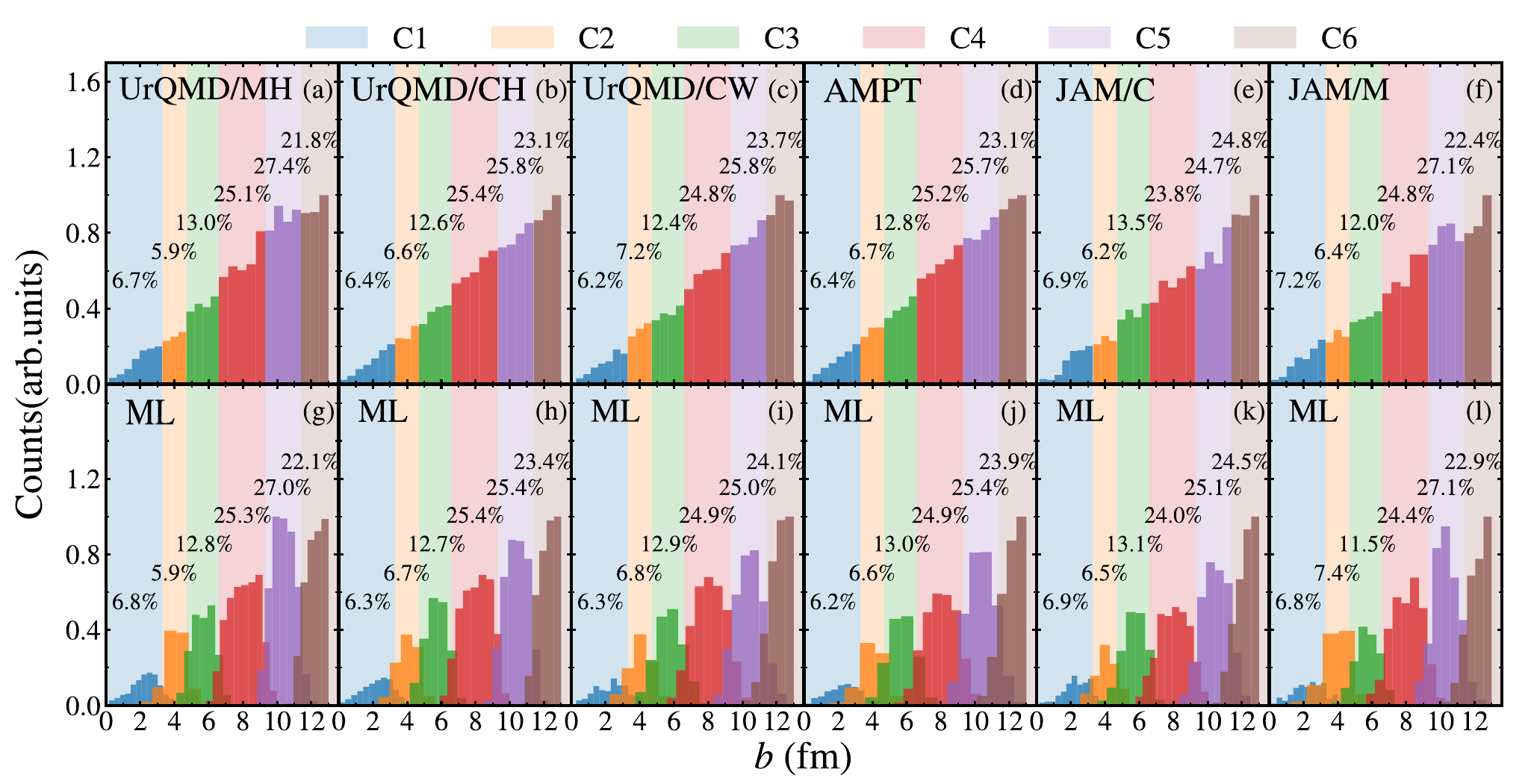}
    \caption{The distributions of impact parameter for Au+Au collision at $\sqrt{s_{NN}}$ = 4 GeV. The upper panels (a–f) show results from transport models (UrQMD in different modes, AMPT, JAM/C, and JAM/M), while the lower panels (g–l) present the corresponding results predicted by ML algorithm. Shaded bands denote centrality classes C1–C6. The percentages in each panel denote the fraction of events in the corresponding centrality class.}
    \label{fig:6}
\end{figure*}

\begin{figure*}[!t]
    \centering
    \includegraphics[width=\linewidth]{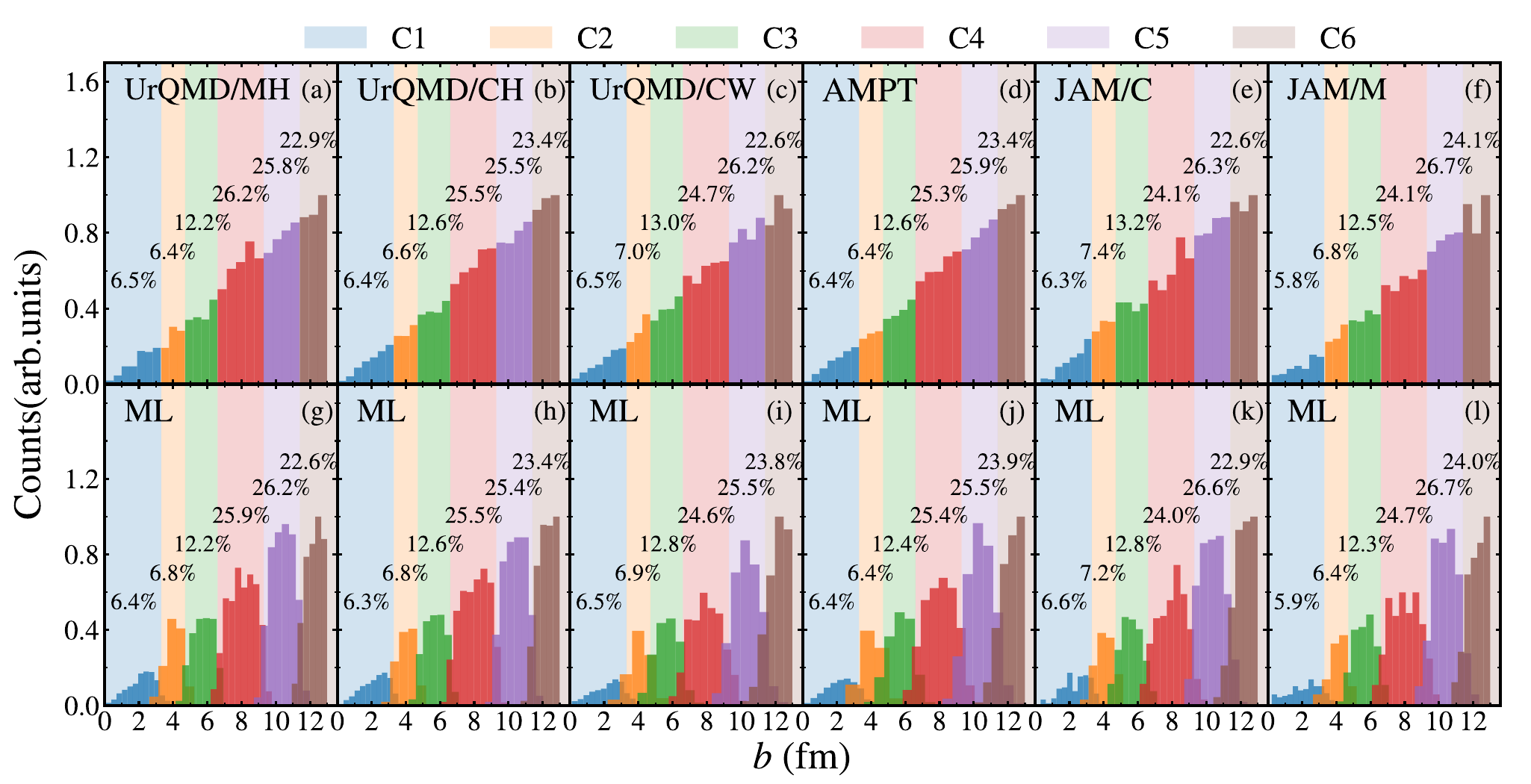}
    \caption{Same as Fig. \ref{fig:6}, but at $\sqrt{s_{NN}}$ = 11 GeV.}
    \label{fig:7}
\end{figure*}

\begin{figure*}[htbp]
    \centering
    \includegraphics[width=\linewidth]{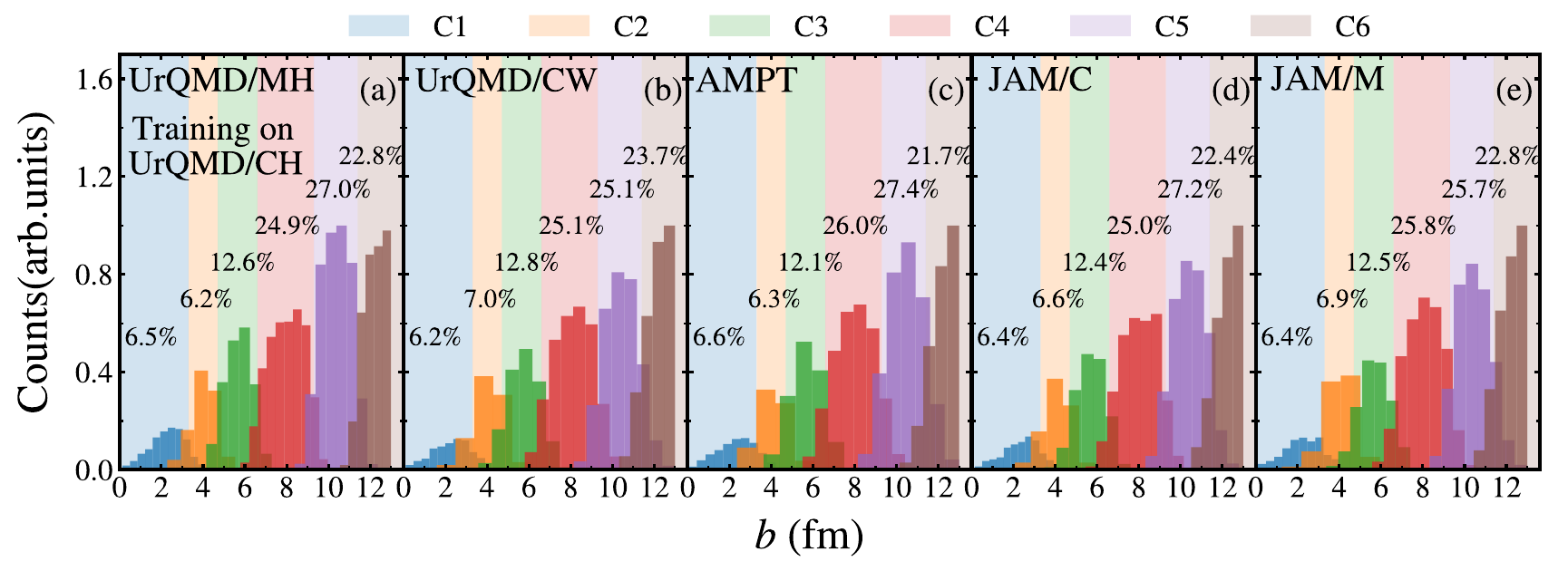}
    \caption{The distribution of the impact parameter for Au+Au collision at $\sqrt{s_{NN}}$ = 4 GeV. The data generated by UrQMD/CH are used to train the ML algorithm, while the test data are generated by (a) UrQMD/MH, (b) UrQMD/CW, (c) AMPT, (d) JAM/C, and (e) JAM/M. Shaded bands denote centrality classes C1–C6, and the percentages indicate the fraction of events in each class.}
    \label{fig:8}
\end{figure*}

\begin{figure*}[htbp]
    \centering
    \includegraphics[width=\linewidth]{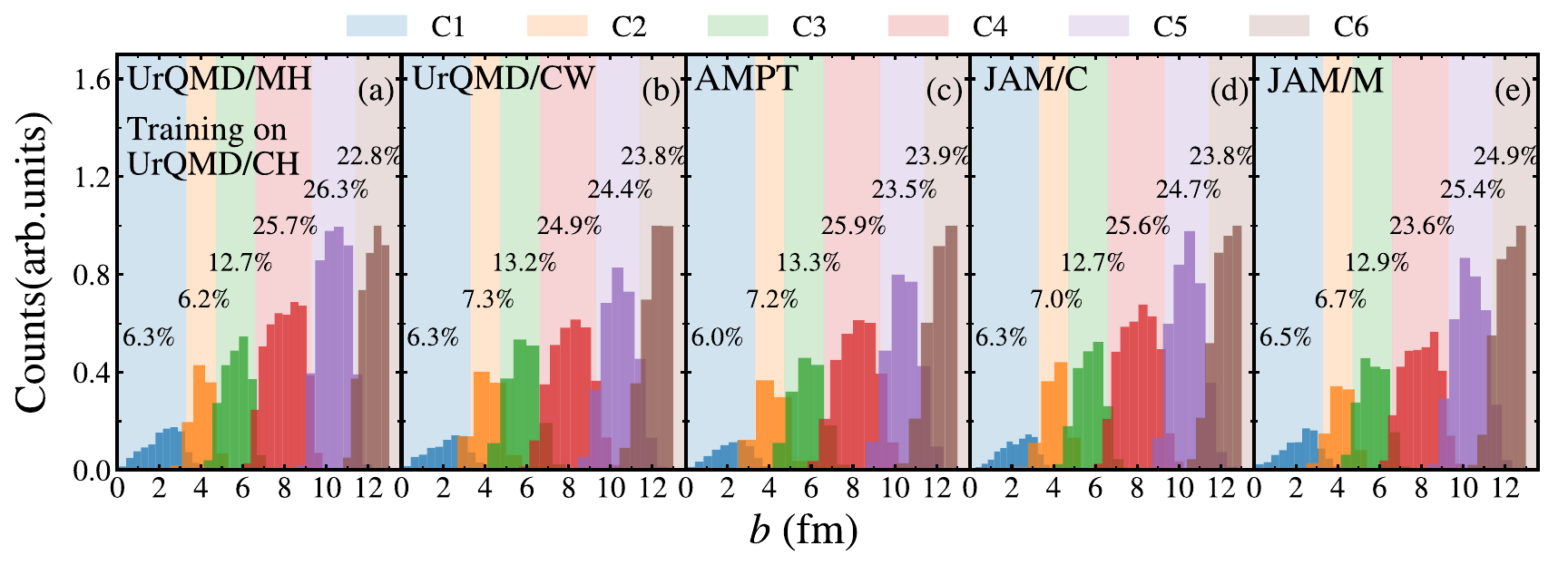}
    \caption{Same as Fig. \ref{fig:8}, but at $\sqrt{s_{NN}}$ = 11 GeV.}
    \label{fig:9}
\end{figure*}

To further demonstrate the performance of the AI approach compared with traditional methods, we compare it with a polynomial-fitting method based on the correlation between the charged-pion yield and the impact parameter. The signed mean error ($\Delta b$) is plotted as a function of the impact parameter in Figs. \ref{fig:4} and \ref{fig:5}. 
In the polynomial-fitting method, the charged-pion yield is used as a centrality-sensitive input feature, and its correlation with the true impact parameter obtained from transport-model simulations is parameterized by a third-order polynomial function,
\begin{equation}
b^{\rm fit}=f(x)=g x^{3}+h x^{2}+i x+j ,
\label{eq:polyfit}
\end{equation}
where $x$ denotes the charged-pion yield, and $g$, $h$, $i$, and $j$ are the fitting parameters determined from the training dataset. The fitted function is then applied to the testing dataset to predict the impact parameter event by event, similar to the conventional benchmark method used in Ref. \cite{OmanaKuttan:2020brq}.
For the ML results, shown by the solid lines, we plot
\(\Delta b=b_i^{\rm pred}-b_i^{\rm true}\), where \(b_i^{\rm pred}\) is predicted by the ML algorithm. And we plot \(\Delta b=b_i^{\rm fit}-b_i^{\rm true}\) for the polynomial-fit results shown by the dashed lines. 

In the top panels of Fig. \ref{fig:4} and Fig. \ref{fig:5}, both the ML training and the polynomial fitting are based on the UrQMD/CH dataset; thus panel (b), where the method is applied to UrQMD/CH itself, serves as the self-consistent reference case. To check the dependence on the chosen reference model, the same procedure is repeated in the bottom panels using AMPT as the training and fitting dataset, for which panel (j) is the corresponding self-consistent case. 
When training and testing data are generated from the same model, the mean error $\Delta b$ approaches zero for both polynomial fitting and ML methods. The ML method exhibits marginally better predictive performance compared to polynomial fitting, consistent with findings in Ref. \cite{OmanaKuttan:2020brq}. When the training and testing data are generated from different models, the $\Delta b$ obtained with ML method is still close to zero while that obtained with the polynomial fitting approach is far from zero. This is understandable because the calculated particle yields vary significantly across different models, and the polynomial parameters fitted based on these yields are not applicable to data obtained from different models. The above results demonstrate the feasibility of training ML algorithms using data generated by transport models and subsequently applying them to predict impact parameter for real experimental data.
We note here that, in the traditional method, only one feature is employed to perform polynomial fitting, as has been widely adopted in many experimental analyses. In principle, multiple features can also be used to establish the complex relationship and then to determine the centrality in the traditional method. In this context, the performance of traditional method may also be improved. 

\subsection{Classification task}

In experimental studies of heavy-ion collisions, each event is categorized into a specific centrality class according to various measurable quantities. In the present work, we classify each event into six classes, as listed in Tab. \ref{class}. LightGBM was utilized to perform six-class classification and cross-validation on the dataset of each model, and the results are presented in Fig. \ref{fig:6} to Fig. \ref{fig:9}.

\begin{table}[htbp]
\caption{\label{different models} Centrality classes in fixed intervals of impact parameter $b_{\rm min}$ - $b_{\rm max}$ for Au + Au collisions. }
\setlength{\tabcolsep}{1.4pt}
\begin{ruledtabular}
\begin{tabular}{ccccc}
 Classes & Centrality & $b_{\rm min}$ & $b_{\rm max}$  \\
\hline
$\rm C1$  & 0-5\%       &0.00               & 3.30   \\
$\rm C2$  & 5-10\%          &3.30              & 4.70     \\
$\rm C3$  & 10-20\%          &4.70               & 6.60     \\
$\rm C4$  & 20-40\%          &6.60               & 9.30      \\
$\rm C5$  & 40-60\%          &9.30               & 11.40      \\
$\rm C6$  & $>$ 60\%          &11.40               &  13.00 \\
\end{tabular}
\end{ruledtabular}\label{class}

\end{table}
Fig. \ref{fig:6} and Fig. \ref{fig:7} systematically compare the impact parameter distributions obtained from six transport models and their corresponding ML predictions at 4 and 11 GeV, respectively. The colored bins labeled C1–C6 explicitly show the proportions of different centrality classes. 
For each centrality class, the corresponding percentages among these six models have small fluctuations, due to differences in the total reaction cross sections among the models.
The ML predicted distributions maintain excellent agreement with the original simulations in their overall trend, and the deviations in the fractions of each centrality class (C1–C6) are controlled at around 1\%. Notably, the ML predicted impact parameter distributions approach smooth, Gaussian-like shapes. This is because the ML model learns the statistical patterns of different impact parameter classes in a data-driven manner, yielding statistically stable continuous distributions. 

Fig. \ref{fig:8} and Fig. \ref{fig:9} show the results of the cross-validation analysis at 4 and 11 GeV, where the ML model is trained on the UrQMD/CH dataset and tested on the set of UrQMD/MH, UrQMD/CW, AMPT, JAM/C and JAM/M as shown in panels (a)–(e). The cross-validated results indicate that the predicted impact parameter distributions remain in close alignment with the original simulations regardless of whether the energy is 4 or 11 GeV, even when trained on a single model and applied to others. This indicates that the ML model has learned the general statistical characteristics of the distribution of impact parameters in HICs, which confirms the robustness of the ML framework and highlights its potential as a reliable tool for rapid and precise inference in nuclear theory research.

\subsection{Clustering task}
\begin{figure*}
    \centering
    \includegraphics[width=\linewidth]{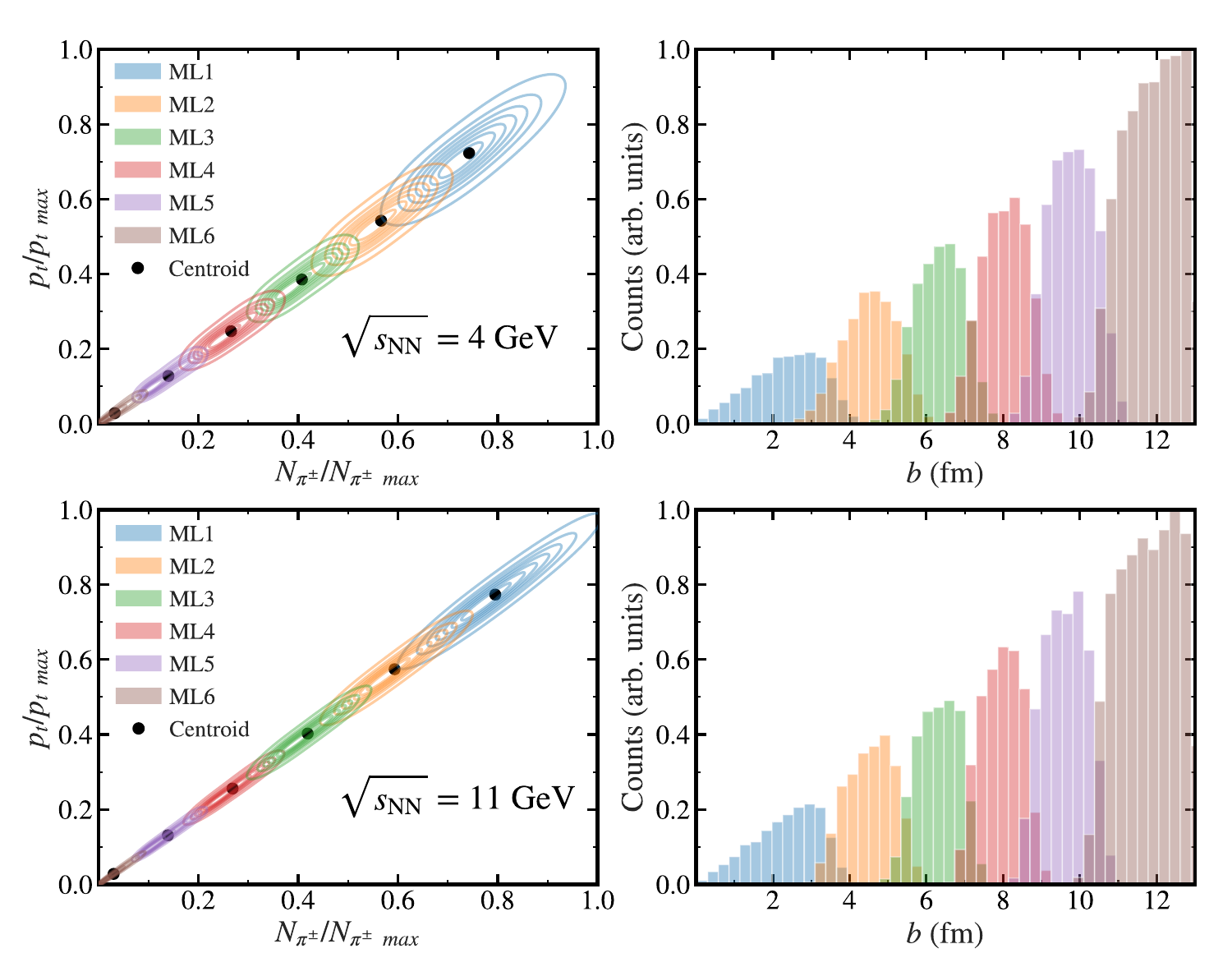}
    \caption{K-means clustering results for Au+Au collisions at $\sqrt{s_{NN}}$ = 4 (top) and 11 (bottom) GeV, based on the data generated by UrQMD/CH. The left panels show the clustering in the $N_{\pi^{\pm}} / N_{\pi^{\pm}}^{\text{max}}$ and $p_t / p_t^{\text{max}}$, where different colors correspond to clusters ML1–ML6, and the markers denote the corresponding centroids. The right panels present the distributions of the impact parameter for each cluster. }
    \label{fig:10}
\end{figure*}

The K-means clustering algorithm is employed as an unsupervised learning tool to partition impact parameters. The algorithm iteratively assigns each event to one of $k$ clusters by minimizing the within-cluster sum of squared distances to the cluster centroids, thereby grouping events with similar statistical properties \cite{LiLI:2022vlp,Oleksiyuk:2024hru,Lobato:2022ukl}. In this work, we divide events with different impact parameters into six clusters via K-means. This unsupervised partitioning not only aligns well with the traditional geometric definition of the impact parameter but also provides a data-driven framework for cross-validating the centrality determination, enhancing the objectivity and reliability of the classification scheme.

Fig. \ref{fig:10} presents the K-means clustering results for Au+Au collisions at $\sqrt{s_{NN}}$ = 4 and 11 GeV (top panel and bottom panel). The left panel displays the 2D distribution of events in the space of normalized multiplicity of charged pions $N_{\pi^{\pm}} / N_{\pi^{\pm}}^{\text{max}}$ (horizontal axis) and normalized transverse momentum $p_t / p_t^{\text{max}}$ (vertical axis), where the K-means algorithm successfully partitions the full event sample into six distinct clusters (ML1–ML6). The contour lines illustrate the event density within each cluster, while the black dots mark the centroid of each group, revealing a clear, linear diagonal trend that reflects the strong positive correlation between multiplicity of charged pions and transverse momentum as a function of collision centrality. The right panel shows the distributions of impact parameter $b$ for each of the six clusters, with filled regions highlighting the event populations for each class. It can be observed that ML1 (blue) corresponds to central collisions with small impact parameters ($b\approx0-4$ fm), characterized by the highest normalized multiplicity of charged pions and transverse momentum, and ML6 (brown) corresponds to peripheral collisions with large impact parameters ($b\approx10-13$ fm), marked by the lowest particle production and transverse momentum. ML2–ML5 (orange, green, red, purple) span the semi-central to mid-peripheral collision regimes, with a smooth, monotonic transition in both observables and impact parameter between the central and peripheral extremes. Critically, the six clusters exhibit minimal overlap in the 2D phase space, whether at 4 or 11 GeV. This confirms that the unsupervised K-means algorithm autonomously identifies the physical centrality structure of heavy-ion collisions directly from experimental-style observables, without relying on prior model-based binning at both energies.

These results demonstrate that K-means clustering provides a robust, data-driven method for centrality determination in heavy-ion collisions. The strong linear correlation between multiplicity of charged pions, transverse momentum, and impact parameter across all clusters confirms that these observables are reliable proxies for centrality, while the clear separation of clusters validates the physical interpretability of the unsupervised classification. Collectively, these findings establish a complete, model-independent framework for centrality calibration, bridging the gap between transport model simulations and experimental data analysis, and highlighting the potential of ML to enhance centrality determination in low-energy heavy-ion collision experiments.

\section{Conclusion and outlook}
\label{Conclusion}
In conclusion, we have systematically investigated the capability of the machine learning (ML) algorithms to reconstruct the impact parameter in Au+Au collisions at $\sqrt{s_{\rm NN}}$ = 4 and 11 GeV, using labeled data generated from three mainstream transport models (UrQMD, AMPT, and JAM) under different initialization setups. The performance of ML algorithms was comprehensively cross-validated through regression, classification, and clustering tasks, and its effectiveness was compared with that of the traditional polynomial fitting method based on yield of charged pions.

In the regression task, LightGBM achieves a mean absolute error (MAE) of 0.2-0.4 fm even when trained on data from one transport model and tested on another, demonstrating strong generalization ability that surpasses the traditional polynomial fitting method, which only works for data from the same model. In the classification task, LightGBM accurately predicts six centrality classes, with deviations in the fraction of each class controlled around 1\% and excellent consistency with the original transport model simulations, even in cross-validation across different models. In the unsupervised clustering task, the K-means algorithm autonomously partitions collision events into six clusters, showing clear separation between central, semi-central, and peripheral collisions without relying on prior model-based binning, thus validating the objectivity and reliability of data-driven centrality determination.

Our study demonstrates that the ML algorithms, trained using data generated from transport models, can reliably and robustly determine the impact parameter in heavy-ion collisions at $\sqrt{s_{\rm NN}}$ = 4 and 11 GeV, with performance significantly superior to that of traditional methods. This ML model’s ability to generalize across different transport models addresses the critical issue of model dependence in centrality determination, laying a solid foundation for its application to real experimental data from facilities such as STAR-FXT, NICA/MPD, and FAIR/CBM. 
For future work, we will incorporate additional experimental observables, such as the yields of light charged particles and particle correlations, obtained from a broader set of event generators, e.g., SMASH, UrQMD-hybrid, GiBUU, and IQMD, as input features to further improve the accuracy of impact-parameter prediction, especially for peripheral collisions where fluctuations are more significant. 
Additionally, we will explore the potential of other advanced ML algorithms, such as transformer-based models recently introduced for latent representation learning of heavy-ion collision events \cite{Zhang:2025maskpoint}, to further enhance the performance of impact parameter reconstruction. Together with existing ML-based impact-parameter reconstruction studies \cite{OmanaKuttan:2020brq,Li:2021plq,Li:2020qqn}, this may contribute to a more comprehensive and reliable framework for centrality determination in heavy-ion collision physics.

\section*{Acknowledgements}
We acknowledge support by the computing server C3S2 Huzhou Normal University. This work is supported in part by the National Natural Science Foundation of China (Nos. 12335008 and 12505143), 
the National Key Research and Development Program of China (No. 2023YFA1606402), 
the Zhejiang Provincial Natural Science Foundation of China (No. LQN25A050003), 
the Huzhou Natural Science Foundation (No. 2024YZ28), 
the Scientific Research Fund of the Zhejiang Provincial Education Department (No. Y202353782),
the Foundation of National Key Laboratory of Plasma Physics (Grant No. 6142A04230203),
the Fund for Shanxi "1331 Project" Key Subjects Construction. 
\bibliographystyle{apsrev4-1}
\bibliography{refer}

\begin{thebibliography}{88}%
\makeatletter
\providecommand \@ifxundefined [1]{%
 \@ifx{#1\undefined}
}%
\providecommand \@ifnum [1]{%
 \ifnum #1\expandafter \@firstoftwo
 \else \expandafter \@secondoftwo
 \fi
}%
\providecommand \@ifx [1]{%
 \ifx #1\expandafter \@firstoftwo
 \else \expandafter \@secondoftwo
 \fi
}%
\providecommand \natexlab [1]{#1}%
\providecommand \enquote  [1]{``#1''}%
\providecommand \bibnamefont  [1]{#1}%
\providecommand \bibfnamefont [1]{#1}%
\providecommand \citenamefont [1]{#1}%
\providecommand \href@noop [0]{\@secondoftwo}%
\providecommand \href [0]{\begingroup \@sanitize@url \@href}%
\providecommand \@href[1]{\@@startlink{#1}\@@href}%
\providecommand \@@href[1]{\endgroup#1\@@endlink}%
\providecommand \@sanitize@url [0]{\catcode `\\12\catcode `\$12\catcode `\&12\catcode `\#12\catcode `\^12\catcode `\_12\catcode `\%12\relax}%
\providecommand \@@startlink[1]{}%
\providecommand \@@endlink[0]{}%
\providecommand \url  [0]{\begingroup\@sanitize@url \@url }%
\providecommand \@url [1]{\endgroup\@href {#1}{\urlprefix }}%
\providecommand \urlprefix  [0]{URL }%
\providecommand \Eprint [0]{\href }%
\providecommand \doibase [0]{http://dx.doi.org/}%
\providecommand \selectlanguage [0]{\@gobble}%
\providecommand \bibinfo  [0]{\@secondoftwo}%
\providecommand \bibfield  [0]{\@secondoftwo}%
\providecommand \translation [1]{[#1]}%
\providecommand \BibitemOpen [0]{}%
\providecommand \bibitemStop [0]{}%
\providecommand \bibitemNoStop [0]{.\EOS\space}%
\providecommand \EOS [0]{\spacefactor3000\relax}%
\providecommand \BibitemShut  [1]{\csname bibitem#1\endcsname}%
\let\auto@bib@innerbib\@empty
\bibitem [{\citenamefont {Danielewicz}\ \emph {et~al.}(2002)\citenamefont {Danielewicz}, \citenamefont {Lacey},\ and\ \citenamefont {Lynch}}]{Danielewicz:2002pu}%
  \BibitemOpen
  \bibfield  {author} {\bibinfo {author} {\bibfnamefont {P.}~\bibnamefont {Danielewicz}}, \bibinfo {author} {\bibfnamefont {R.}~\bibnamefont {Lacey}}, \ and\ \bibinfo {author} {\bibfnamefont {W.~G.}\ \bibnamefont {Lynch}},\ }\href {\doibase 10.1126/science.1078070} {\bibfield  {journal} {\bibinfo  {journal} {Science}\ }\textbf {\bibinfo {volume} {298}},\ \bibinfo {pages} {1592} (\bibinfo {year} {2002})},\ \Eprint {http://arxiv.org/abs/nucl-th/0208016} {arXiv:nucl-th/0208016} \BibitemShut {NoStop}%
\bibitem [{\citenamefont {Fukushima}\ and\ \citenamefont {Hatsuda}(2011)}]{Fukushima:2010bq}%
  \BibitemOpen
  \bibfield  {author} {\bibinfo {author} {\bibfnamefont {K.}~\bibnamefont {Fukushima}}\ and\ \bibinfo {author} {\bibfnamefont {T.}~\bibnamefont {Hatsuda}},\ }\href {\doibase 10.1088/0034-4885/74/1/014001} {\bibfield  {journal} {\bibinfo  {journal} {Rept. Prog. Phys.}\ }\textbf {\bibinfo {volume} {74}},\ \bibinfo {pages} {014001} (\bibinfo {year} {2011})},\ \Eprint {http://arxiv.org/abs/1005.4814} {arXiv:1005.4814 [hep-ph]} \BibitemShut {NoStop}%
\bibitem [{\citenamefont {Sorensen}\ \emph {et~al.}(2024)\citenamefont {Sorensen} \emph {et~al.}}]{Sorensen:2023zkk}%
  \BibitemOpen
  \bibfield  {author} {\bibinfo {author} {\bibfnamefont {A.}~\bibnamefont {Sorensen}} \emph {et~al.},\ }\href {\doibase 10.1016/j.ppnp.2023.104080} {\bibfield  {journal} {\bibinfo  {journal} {Prog. Part. Nucl. Phys.}\ }\textbf {\bibinfo {volume} {134}},\ \bibinfo {pages} {104080} (\bibinfo {year} {2024})},\ \Eprint {http://arxiv.org/abs/2301.13253} {arXiv:2301.13253 [nucl-th]} \BibitemShut {NoStop}%
\bibitem [{\citenamefont {Guo}\ \emph {et~al.}(2024)\citenamefont {Guo} \emph {et~al.}}]{Guo:2024zij}%
  \BibitemOpen
  \bibfield  {author} {\bibinfo {author} {\bibfnamefont {D.}~\bibnamefont {Guo}} \emph {et~al.},\ }\href {\doibase 10.1140/epja/s10050-024-01245-2} {\bibfield  {journal} {\bibinfo  {journal} {Eur. Phys. J. A}\ }\textbf {\bibinfo {volume} {60}},\ \bibinfo {pages} {36} (\bibinfo {year} {2024})}\BibitemShut {NoStop}%
\bibitem [{\citenamefont {Klochkov}\ \emph {et~al.}(2021)\citenamefont {Klochkov}, \citenamefont {Collaboration} \emph {et~al.}}]{Klochkov:2021eyo}%
  \BibitemOpen
  \bibfield  {author} {\bibinfo {author} {\bibfnamefont {V.}~\bibnamefont {Klochkov}}, \bibinfo {author} {\bibfnamefont {C.}~\bibnamefont {Collaboration}},  \emph {et~al.},\ }\href {\doibase 10.1016/j.nuclphysa.2020.121945} {\bibfield  {journal} {\bibinfo  {journal} {Nuclear Physics A}\ }\textbf {\bibinfo {volume} {1005}},\ \bibinfo {pages} {121945} (\bibinfo {year} {2021})}\BibitemShut {NoStop}%
\bibitem [{\citenamefont {Adamczyk}\ \emph {et~al.}(2017)\citenamefont {Adamczyk} \emph {et~al.}}]{STAR:2017sal}%
  \BibitemOpen
  \bibfield  {author} {\bibinfo {author} {\bibfnamefont {L.}~\bibnamefont {Adamczyk}} \emph {et~al.} (\bibinfo {collaboration} {STAR}),\ }\href {\doibase 10.1103/PhysRevC.96.044904} {\bibfield  {journal} {\bibinfo  {journal} {Phys. Rev. C}\ }\textbf {\bibinfo {volume} {96}},\ \bibinfo {pages} {044904} (\bibinfo {year} {2017})},\ \Eprint {http://arxiv.org/abs/1701.07065} {arXiv:1701.07065 [nucl-ex]} \BibitemShut {NoStop}%
\bibitem [{\citenamefont {Abgaryan}\ \emph {et~al.}(2022)\citenamefont {Abgaryan} \emph {et~al.}}]{MPD:2022qhn}%
  \BibitemOpen
  \bibfield  {author} {\bibinfo {author} {\bibfnamefont {V.}~\bibnamefont {Abgaryan}} \emph {et~al.} (\bibinfo {collaboration} {MPD}),\ }\href {\doibase 10.1140/epja/s10050-022-00750-6} {\bibfield  {journal} {\bibinfo  {journal} {Eur. Phys. J. A}\ }\textbf {\bibinfo {volume} {58}},\ \bibinfo {pages} {140} (\bibinfo {year} {2022})},\ \Eprint {http://arxiv.org/abs/2202.08970} {arXiv:2202.08970 [physics.ins-det]} \BibitemShut {NoStop}%
\bibitem [{\citenamefont {Syresin}\ \emph {et~al.}(2025)\citenamefont {Syresin} \emph {et~al.}}]{Syresin:2025lye}%
  \BibitemOpen
  \bibfield  {author} {\bibinfo {author} {\bibfnamefont {E.}~\bibnamefont {Syresin}} \emph {et~al.},\ }\href {\doibase 10.1088/1674-1137/adbad0} {\bibfield  {journal} {\bibinfo  {journal} {Chin. Phys. C}\ }\textbf {\bibinfo {volume} {49}},\ \bibinfo {pages} {074003} (\bibinfo {year} {2025})}\BibitemShut {NoStop}%
\bibitem [{\citenamefont {Zhu}\ \emph {et~al.}(2025)\citenamefont {Zhu}, \citenamefont {Wu},\ and\ \citenamefont {Qin}}]{Zhu:2024tns}%
  \BibitemOpen
  \bibfield  {author} {\bibinfo {author} {\bibfnamefont {J.}~\bibnamefont {Zhu}}, \bibinfo {author} {\bibfnamefont {X.-Y.}\ \bibnamefont {Wu}}, \ and\ \bibinfo {author} {\bibfnamefont {G.-Y.}\ \bibnamefont {Qin}},\ }\href {\doibase 10.1088/1674-1137/ada7d1} {\bibfield  {journal} {\bibinfo  {journal} {Chin. Phys. C}\ }\textbf {\bibinfo {volume} {49}},\ \bibinfo {pages} {044103} (\bibinfo {year} {2025})},\ \Eprint {http://arxiv.org/abs/2401.15536} {arXiv:2401.15536 [hep-ph]} \BibitemShut {NoStop}%
\bibitem [{\citenamefont {Wolter}\ \emph {et~al.}(2022)\citenamefont {Wolter} \emph {et~al.}}]{TMEP:2022xjg}%
  \BibitemOpen
  \bibfield  {author} {\bibinfo {author} {\bibfnamefont {H.}~\bibnamefont {Wolter}} \emph {et~al.} (\bibinfo {collaboration} {TMEP}),\ }\href {\doibase 10.1016/j.ppnp.2022.103962} {\bibfield  {journal} {\bibinfo  {journal} {Prog. Part. Nucl. Phys.}\ }\textbf {\bibinfo {volume} {125}},\ \bibinfo {pages} {103962} (\bibinfo {year} {2022})},\ \Eprint {http://arxiv.org/abs/2202.06672} {arXiv:2202.06672 [nucl-th]} \BibitemShut {NoStop}%
\bibitem [{\citenamefont {Adamczyk}\ \emph {et~al.}(2014)\citenamefont {Adamczyk} \emph {et~al.}}]{STAR:2013gus}%
  \BibitemOpen
  \bibfield  {author} {\bibinfo {author} {\bibfnamefont {L.}~\bibnamefont {Adamczyk}} \emph {et~al.} (\bibinfo {collaboration} {STAR}),\ }\href {\doibase 10.1103/PhysRevLett.112.032302} {\bibfield  {journal} {\bibinfo  {journal} {Phys. Rev. Lett.}\ }\textbf {\bibinfo {volume} {112}},\ \bibinfo {pages} {032302} (\bibinfo {year} {2014})},\ \Eprint {http://arxiv.org/abs/1309.5681} {arXiv:1309.5681 [nucl-ex]} \BibitemShut {NoStop}%
\bibitem [{\citenamefont {Zhang}\ \emph {et~al.}(2026)\citenamefont {Zhang}, \citenamefont {Zhang}, \citenamefont {Luo},\ and\ \citenamefont {Xu}}]{Zhang:2025ale}%
  \BibitemOpen
  \bibfield  {author} {\bibinfo {author} {\bibfnamefont {X.}~\bibnamefont {Zhang}}, \bibinfo {author} {\bibfnamefont {Y.}~\bibnamefont {Zhang}}, \bibinfo {author} {\bibfnamefont {X.}~\bibnamefont {Luo}}, \ and\ \bibinfo {author} {\bibfnamefont {N.}~\bibnamefont {Xu}},\ }\href {\doibase 10.1088/1674-1137/ae0995} {\bibfield  {journal} {\bibinfo  {journal} {Chin. Phys. C}\ }\textbf {\bibinfo {volume} {50}},\ \bibinfo {pages} {011003} (\bibinfo {year} {2026})},\ \Eprint {http://arxiv.org/abs/2506.18832} {arXiv:2506.18832 [nucl-ex]} \BibitemShut {NoStop}%
\bibitem [{\citenamefont {Cavata}\ \emph {et~al.}(1990)\citenamefont {Cavata}, \citenamefont {Demoulins}, \citenamefont {Gosset}, \citenamefont {Lemaire}, \citenamefont {L'Hote}, \citenamefont {Poitou},\ and\ \citenamefont {Valette}}]{Cavata:1990gk}%
  \BibitemOpen
  \bibfield  {author} {\bibinfo {author} {\bibfnamefont {C.}~\bibnamefont {Cavata}}, \bibinfo {author} {\bibfnamefont {M.}~\bibnamefont {Demoulins}}, \bibinfo {author} {\bibfnamefont {J.}~\bibnamefont {Gosset}}, \bibinfo {author} {\bibfnamefont {M.~C.}\ \bibnamefont {Lemaire}}, \bibinfo {author} {\bibfnamefont {D.}~\bibnamefont {L'Hote}}, \bibinfo {author} {\bibfnamefont {J.}~\bibnamefont {Poitou}}, \ and\ \bibinfo {author} {\bibfnamefont {O.}~\bibnamefont {Valette}},\ }\href {\doibase 10.1103/PhysRevC.42.1760} {\bibfield  {journal} {\bibinfo  {journal} {Phys. Rev. C}\ }\textbf {\bibinfo {volume} {42}},\ \bibinfo {pages} {1760} (\bibinfo {year} {1990})}\BibitemShut {NoStop}%
\bibitem [{\citenamefont {Miller}\ \emph {et~al.}(2007)\citenamefont {Miller}, \citenamefont {Reygers}, \citenamefont {Sanders},\ and\ \citenamefont {Steinberg}}]{Miller:2007ri}%
  \BibitemOpen
  \bibfield  {author} {\bibinfo {author} {\bibfnamefont {M.~L.}\ \bibnamefont {Miller}}, \bibinfo {author} {\bibfnamefont {K.}~\bibnamefont {Reygers}}, \bibinfo {author} {\bibfnamefont {S.~J.}\ \bibnamefont {Sanders}}, \ and\ \bibinfo {author} {\bibfnamefont {P.}~\bibnamefont {Steinberg}},\ }\href {\doibase 10.1146/annurev.nucl.57.090506.123020} {\bibfield  {journal} {\bibinfo  {journal} {Ann. Rev. Nucl. Part. Sci.}\ }\textbf {\bibinfo {volume} {57}},\ \bibinfo {pages} {205} (\bibinfo {year} {2007})},\ \Eprint {http://arxiv.org/abs/nucl-ex/0701025} {arXiv:nucl-ex/0701025} \BibitemShut {NoStop}%
\bibitem [{\citenamefont {Adamczewski-Musch}\ \emph {et~al.}(2018)\citenamefont {Adamczewski-Musch} \emph {et~al.}}]{HADES:2017def}%
  \BibitemOpen
  \bibfield  {author} {\bibinfo {author} {\bibfnamefont {J.}~\bibnamefont {Adamczewski-Musch}} \emph {et~al.} (\bibinfo {collaboration} {HADES}),\ }\href {\doibase 10.1140/epja/i2018-12513-7} {\bibfield  {journal} {\bibinfo  {journal} {Eur. Phys. J. A}\ }\textbf {\bibinfo {volume} {54}},\ \bibinfo {pages} {85} (\bibinfo {year} {2018})},\ \Eprint {http://arxiv.org/abs/1712.07993} {arXiv:1712.07993 [nucl-ex]} \BibitemShut {NoStop}%
\bibitem [{\citenamefont {Abelev}\ \emph {et~al.}(2010)\citenamefont {Abelev} \emph {et~al.}}]{STAR:2009sxc}%
  \BibitemOpen
  \bibfield  {author} {\bibinfo {author} {\bibfnamefont {B.~I.}\ \bibnamefont {Abelev}} \emph {et~al.} (\bibinfo {collaboration} {STAR}),\ }\href {\doibase 10.1103/PhysRevC.81.024911} {\bibfield  {journal} {\bibinfo  {journal} {Phys. Rev. C}\ }\textbf {\bibinfo {volume} {81}},\ \bibinfo {pages} {024911} (\bibinfo {year} {2010})},\ \Eprint {http://arxiv.org/abs/0909.4131} {arXiv:0909.4131 [nucl-ex]} \BibitemShut {NoStop}%
\bibitem [{\citenamefont {Adamczyk}\ \emph {et~al.}(2012)\citenamefont {Adamczyk} \emph {et~al.}}]{STAR:2012och}%
  \BibitemOpen
  \bibfield  {author} {\bibinfo {author} {\bibfnamefont {L.}~\bibnamefont {Adamczyk}} \emph {et~al.} (\bibinfo {collaboration} {STAR}),\ }\href {\doibase 10.1103/PhysRevC.86.054908} {\bibfield  {journal} {\bibinfo  {journal} {Phys. Rev. C}\ }\textbf {\bibinfo {volume} {86}},\ \bibinfo {pages} {054908} (\bibinfo {year} {2012})},\ \Eprint {http://arxiv.org/abs/1206.5528} {arXiv:1206.5528 [nucl-ex]} \BibitemShut {NoStop}%
\bibitem [{\citenamefont {Adam}\ \emph {et~al.}(2020{\natexlab{a}})\citenamefont {Adam} \emph {et~al.}}]{STAR:2019vcp}%
  \BibitemOpen
  \bibfield  {author} {\bibinfo {author} {\bibfnamefont {J.}~\bibnamefont {Adam}} \emph {et~al.} (\bibinfo {collaboration} {STAR}),\ }\href {\doibase 10.1103/PhysRevC.101.024905} {\bibfield  {journal} {\bibinfo  {journal} {Phys. Rev. C}\ }\textbf {\bibinfo {volume} {101}},\ \bibinfo {pages} {024905} (\bibinfo {year} {2020}{\natexlab{a}})},\ \Eprint {http://arxiv.org/abs/1908.03585} {arXiv:1908.03585 [nucl-ex]} \BibitemShut {NoStop}%
\bibitem [{\citenamefont {Adam}\ \emph {et~al.}(2020{\natexlab{b}})\citenamefont {Adam} \emph {et~al.}}]{STAR:2019bjj}%
  \BibitemOpen
  \bibfield  {author} {\bibinfo {author} {\bibfnamefont {J.}~\bibnamefont {Adam}} \emph {et~al.} (\bibinfo {collaboration} {STAR}),\ }\href {\doibase 10.1103/PhysRevC.102.034909} {\bibfield  {journal} {\bibinfo  {journal} {Phys. Rev. C}\ }\textbf {\bibinfo {volume} {102}},\ \bibinfo {pages} {034909} (\bibinfo {year} {2020}{\natexlab{b}})},\ \Eprint {http://arxiv.org/abs/1906.03732} {arXiv:1906.03732 [nucl-ex]} \BibitemShut {NoStop}%
\bibitem [{\citenamefont {Abdallah}\ \emph {et~al.}(2023)\citenamefont {Abdallah} \emph {et~al.}}]{STAR:2022etb}%
  \BibitemOpen
  \bibfield  {author} {\bibinfo {author} {\bibfnamefont {M.}~\bibnamefont {Abdallah}} \emph {et~al.} (\bibinfo {collaboration} {STAR}),\ }\href {\doibase 10.1103/PhysRevC.107.024908} {\bibfield  {journal} {\bibinfo  {journal} {Phys. Rev. C}\ }\textbf {\bibinfo {volume} {107}},\ \bibinfo {pages} {024908} (\bibinfo {year} {2023})},\ \Eprint {http://arxiv.org/abs/2209.11940} {arXiv:2209.11940 [nucl-ex]} \BibitemShut {NoStop}%
\bibitem [{\citenamefont {Yue}\ \emph {et~al.}(2026)\citenamefont {Yue}, \citenamefont {Li}, \citenamefont {Wang}, \citenamefont {Li},\ and\ \citenamefont {Liu}}]{Yue:2026flb}%
  \BibitemOpen
  \bibfield  {author} {\bibinfo {author} {\bibfnamefont {X.}~\bibnamefont {Yue}}, \bibinfo {author} {\bibfnamefont {P.}~\bibnamefont {Li}}, \bibinfo {author} {\bibfnamefont {Y.}~\bibnamefont {Wang}}, \bibinfo {author} {\bibfnamefont {Q.}~\bibnamefont {Li}}, \ and\ \bibinfo {author} {\bibfnamefont {F.}~\bibnamefont {Liu}},\ }\href {\doibase 10.1103/k2vx-4898} {\bibfield  {journal} {\bibinfo  {journal} {Phys. Rev. C}\ }\textbf {\bibinfo {volume} {113}},\ \bibinfo {pages} {024905} (\bibinfo {year} {2026})},\ \Eprint {http://arxiv.org/abs/2601.20491} {arXiv:2601.20491 [nucl-th]} \BibitemShut {NoStop}%
\bibitem [{\citenamefont {Lukasik}\ \emph {et~al.}(1997)\citenamefont {Lukasik} \emph {et~al.}}]{Lukasik:1997zz}%
  \BibitemOpen
  \bibfield  {author} {\bibinfo {author} {\bibfnamefont {J.}~\bibnamefont {Lukasik}} \emph {et~al.},\ }\href {\doibase 10.1103/PhysRevC.55.1906} {\bibfield  {journal} {\bibinfo  {journal} {Phys. Rev. C}\ }\textbf {\bibinfo {volume} {55}},\ \bibinfo {pages} {1906} (\bibinfo {year} {1997})}\BibitemShut {NoStop}%
\bibitem [{\citenamefont {Lukasik}\ \emph {et~al.}(2005)\citenamefont {Lukasik} \emph {et~al.}}]{Lukasik:2004df}%
  \BibitemOpen
  \bibfield  {author} {\bibinfo {author} {\bibfnamefont {J.}~\bibnamefont {Lukasik}} \emph {et~al.},\ }\href {\doibase 10.1016/j.physletb.2004.12.076} {\bibfield  {journal} {\bibinfo  {journal} {Phys. Lett. B}\ }\textbf {\bibinfo {volume} {608}},\ \bibinfo {pages} {223} (\bibinfo {year} {2005})},\ \Eprint {http://arxiv.org/abs/nucl-ex/0410030} {arXiv:nucl-ex/0410030} \BibitemShut {NoStop}%
\bibitem [{\citenamefont {Sch{\"u}ttauf}\ \emph {et~al.}(1996)\citenamefont {Sch{\"u}ttauf} \emph {et~al.}}]{Schuttauf:1996ci}%
  \BibitemOpen
  \bibfield  {author} {\bibinfo {author} {\bibfnamefont {A.}~\bibnamefont {Sch{\"u}ttauf}} \emph {et~al.},\ }\href {\doibase 10.1016/0375-9474(96)00239-4} {\bibfield  {journal} {\bibinfo  {journal} {Nucl. Phys. A}\ }\textbf {\bibinfo {volume} {607}},\ \bibinfo {pages} {457} (\bibinfo {year} {1996})},\ \Eprint {http://arxiv.org/abs/nucl-ex/9606001} {arXiv:nucl-ex/9606001} \BibitemShut {NoStop}%
\bibitem [{\citenamefont {Reisdorf}\ \emph {et~al.}(1997)\citenamefont {Reisdorf} \emph {et~al.}}]{FOPI:1996cjz}%
  \BibitemOpen
  \bibfield  {author} {\bibinfo {author} {\bibfnamefont {W.}~\bibnamefont {Reisdorf}} \emph {et~al.} (\bibinfo {collaboration} {FOPI}),\ }\href {\doibase 10.1016/S0375-9474(96)00388-0} {\bibfield  {journal} {\bibinfo  {journal} {Nucl. Phys. A}\ }\textbf {\bibinfo {volume} {612}},\ \bibinfo {pages} {493} (\bibinfo {year} {1997})},\ \Eprint {http://arxiv.org/abs/nucl-ex/9610009} {arXiv:nucl-ex/9610009} \BibitemShut {NoStop}%
\bibitem [{\citenamefont {Le~F{\`e}vre}\ \emph {et~al.}(2016)\citenamefont {Le~F{\`e}vre}, \citenamefont {Leifels}, \citenamefont {Reisdorf}, \citenamefont {Aichelin},\ and\ \citenamefont {Hartnack}}]{LeFevre:2015paj}%
  \BibitemOpen
  \bibfield  {author} {\bibinfo {author} {\bibfnamefont {A.}~\bibnamefont {Le~F{\`e}vre}}, \bibinfo {author} {\bibfnamefont {Y.}~\bibnamefont {Leifels}}, \bibinfo {author} {\bibfnamefont {W.}~\bibnamefont {Reisdorf}}, \bibinfo {author} {\bibfnamefont {J.}~\bibnamefont {Aichelin}}, \ and\ \bibinfo {author} {\bibfnamefont {C.}~\bibnamefont {Hartnack}},\ }\href {\doibase 10.1016/j.nuclphysa.2015.09.015} {\bibfield  {journal} {\bibinfo  {journal} {Nucl. Phys. A}\ }\textbf {\bibinfo {volume} {945}},\ \bibinfo {pages} {112} (\bibinfo {year} {2016})},\ \Eprint {http://arxiv.org/abs/1501.05246} {arXiv:1501.05246 [nucl-ex]} \BibitemShut {NoStop}%
\bibitem [{\citenamefont {Luo}\ and\ \citenamefont {Xu}(2017)}]{Luo:2017faz}%
  \BibitemOpen
  \bibfield  {author} {\bibinfo {author} {\bibfnamefont {X.}~\bibnamefont {Luo}}\ and\ \bibinfo {author} {\bibfnamefont {N.}~\bibnamefont {Xu}},\ }\href {\doibase 10.1007/s41365-017-0257-0} {\bibfield  {journal} {\bibinfo  {journal} {Nucl. Sci. Tech.}\ }\textbf {\bibinfo {volume} {28}},\ \bibinfo {pages} {112} (\bibinfo {year} {2017})},\ \Eprint {http://arxiv.org/abs/1701.02105} {arXiv:1701.02105 [nucl-ex]} \BibitemShut {NoStop}%
\bibitem [{\citenamefont {Chen}()}]{Lbchen}%
  \BibitemOpen
  \bibfield  {author} {\bibinfo {author} {\bibfnamefont {L.}~\bibnamefont {Chen}},\ }\href {https://indico.cern.ch/event/1438306/contributions/} {\enquote {\bibinfo {title} {Charged pion production in au+au collision at $\sqrt{s_{NN}}$ = 3.2-4.5 gev with the star detector},}\ }\bibinfo {howpublished} {Conference report}\BibitemShut {NoStop}%
\bibitem [{\citenamefont {Werner}\ \emph {et~al.}(2025)\citenamefont {Werner}, \citenamefont {Jahan}, \citenamefont {Karpenko}, \citenamefont {Pierog}, \citenamefont {Stefaniak},\ and\ \citenamefont {Vintache}}]{Werner:2024ntd}%
  \BibitemOpen
  \bibfield  {author} {\bibinfo {author} {\bibfnamefont {K.}~\bibnamefont {Werner}}, \bibinfo {author} {\bibfnamefont {J.}~\bibnamefont {Jahan}}, \bibinfo {author} {\bibfnamefont {I.}~\bibnamefont {Karpenko}}, \bibinfo {author} {\bibfnamefont {T.}~\bibnamefont {Pierog}}, \bibinfo {author} {\bibfnamefont {M.}~\bibnamefont {Stefaniak}}, \ and\ \bibinfo {author} {\bibfnamefont {D.}~\bibnamefont {Vintache}},\ }\href {\doibase 10.1103/PhysRevC.111.014903} {\bibfield  {journal} {\bibinfo  {journal} {Phys. Rev. C}\ }\textbf {\bibinfo {volume} {111}},\ \bibinfo {pages} {014903} (\bibinfo {year} {2025})},\ \Eprint {http://arxiv.org/abs/2401.11275} {arXiv:2401.11275 [hep-ph]} \BibitemShut {NoStop}%
\bibitem [{\citenamefont {Qu}\ \emph {et~al.}(2025)\citenamefont {Qu}, \citenamefont {Zhang},\ and\ \citenamefont {Bao}}]{qu2025nuclear}%
  \BibitemOpen
  \bibfield  {author} {\bibinfo {author} {\bibfnamefont {S.}~\bibnamefont {Qu}}, \bibinfo {author} {\bibfnamefont {J.-Y.}\ \bibnamefont {Zhang}}, \ and\ \bibinfo {author} {\bibfnamefont {M.}~\bibnamefont {Bao}},\ }\href {\doibase 10.1088/1674-1137/ade958} {\bibfield  {journal} {\bibinfo  {journal} {Chin. Phys. C}\ }\textbf {\bibinfo {volume} {49}},\ \bibinfo {pages} {104106} (\bibinfo {year} {2025})}\BibitemShut {NoStop}%
\bibitem [{\citenamefont {Mumpower}\ \emph {et~al.}(2022)\citenamefont {Mumpower}, \citenamefont {Sprouse}, \citenamefont {Lovell},\ and\ \citenamefont {Mohan}}]{Mumpower:2022peg}%
  \BibitemOpen
  \bibfield  {author} {\bibinfo {author} {\bibfnamefont {M.~R.}\ \bibnamefont {Mumpower}}, \bibinfo {author} {\bibfnamefont {T.~M.}\ \bibnamefont {Sprouse}}, \bibinfo {author} {\bibfnamefont {A.~E.}\ \bibnamefont {Lovell}}, \ and\ \bibinfo {author} {\bibfnamefont {A.~T.}\ \bibnamefont {Mohan}},\ }\href {\doibase 10.1103/PhysRevC.106.L021301} {\bibfield  {journal} {\bibinfo  {journal} {Phys. Rev. C}\ }\textbf {\bibinfo {volume} {106}},\ \bibinfo {pages} {L021301} (\bibinfo {year} {2022})},\ \Eprint {http://arxiv.org/abs/2203.10594} {arXiv:2203.10594 [nucl-th]} \BibitemShut {NoStop}%
\bibitem [{\citenamefont {Gao}\ \emph {et~al.}(2021)\citenamefont {Gao}, \citenamefont {Wang}, \citenamefont {L{\"u}}, \citenamefont {Li}, \citenamefont {Shen},\ and\ \citenamefont {Liu}}]{gao2021machine}%
  \BibitemOpen
  \bibfield  {author} {\bibinfo {author} {\bibfnamefont {Z.}~\bibnamefont {Gao}}, \bibinfo {author} {\bibfnamefont {Y.}~\bibnamefont {Wang}}, \bibinfo {author} {\bibfnamefont {H.}~\bibnamefont {L{\"u}}}, \bibinfo {author} {\bibfnamefont {Q.}~\bibnamefont {Li}}, \bibinfo {author} {\bibfnamefont {C.}~\bibnamefont {Shen}}, \ and\ \bibinfo {author} {\bibfnamefont {L.}~\bibnamefont {Liu}},\ }\href {\doibase 10.1007/s41365-021-00958-z} {\bibfield  {journal} {\bibinfo  {journal} {Nuclear Science and Techniques}\ }\textbf {\bibinfo {volume} {32}},\ \bibinfo {pages} {109} (\bibinfo {year} {2021})}\BibitemShut {NoStop}%
\bibitem [{\citenamefont {Y{\"u}ksel}\ \emph {et~al.}(2024)\citenamefont {Y{\"u}ksel}, \citenamefont {Soydaner},\ and\ \citenamefont {Bahtiyar}}]{Yuksel:2024zky}%
  \BibitemOpen
  \bibfield  {author} {\bibinfo {author} {\bibfnamefont {E.}~\bibnamefont {Y{\"u}ksel}}, \bibinfo {author} {\bibfnamefont {D.}~\bibnamefont {Soydaner}}, \ and\ \bibinfo {author} {\bibfnamefont {H.}~\bibnamefont {Bahtiyar}},\ }\href {\doibase 10.1103/PhysRevC.109.064322} {\bibfield  {journal} {\bibinfo  {journal} {Phys. Rev. C}\ }\textbf {\bibinfo {volume} {109}},\ \bibinfo {pages} {064322} (\bibinfo {year} {2024})},\ \Eprint {http://arxiv.org/abs/2401.02824} {arXiv:2401.02824 [nucl-th]} \BibitemShut {NoStop}%
\bibitem [{\citenamefont {Kvasiuk}\ \emph {et~al.}(2020)\citenamefont {Kvasiuk}, \citenamefont {Zabrodin}, \citenamefont {Bravina}, \citenamefont {Didur},\ and\ \citenamefont {Frolov}}]{Kvasiuk:2020izb}%
  \BibitemOpen
  \bibfield  {author} {\bibinfo {author} {\bibfnamefont {Y.}~\bibnamefont {Kvasiuk}}, \bibinfo {author} {\bibfnamefont {E.}~\bibnamefont {Zabrodin}}, \bibinfo {author} {\bibfnamefont {L.}~\bibnamefont {Bravina}}, \bibinfo {author} {\bibfnamefont {I.}~\bibnamefont {Didur}}, \ and\ \bibinfo {author} {\bibfnamefont {M.}~\bibnamefont {Frolov}},\ }\href {\doibase 10.1007/JHEP07(2020)133} {\bibfield  {journal} {\bibinfo  {journal} {JHEP}\ }\textbf {\bibinfo {volume} {07}},\ \bibinfo {pages} {133} (\bibinfo {year} {2020})},\ \Eprint {http://arxiv.org/abs/2004.14409} {arXiv:2004.14409 [nucl-th]} \BibitemShut {NoStop}%
\bibitem [{\citenamefont {Wang}\ and\ \citenamefont {Li}(2023)}]{Wang:2023kcg}%
  \BibitemOpen
  \bibfield  {author} {\bibinfo {author} {\bibfnamefont {Y.}~\bibnamefont {Wang}}\ and\ \bibinfo {author} {\bibfnamefont {Q.}~\bibnamefont {Li}},\ }\href {\doibase 10.1007/s11467-023-1313-3} {\bibfield  {journal} {\bibinfo  {journal} {Front. Phys. (Beijing)}\ }\textbf {\bibinfo {volume} {18}},\ \bibinfo {pages} {64402} (\bibinfo {year} {2023})},\ \Eprint {http://arxiv.org/abs/2305.16686} {arXiv:2305.16686 [nucl-th]} \BibitemShut {NoStop}%
\bibitem [{\citenamefont {Patra}\ \emph {et~al.}(2025)\citenamefont {Patra}, \citenamefont {Malik}, \citenamefont {Pais}, \citenamefont {Zhou}, \citenamefont {Agrawal},\ and\ \citenamefont {Provid{\^e}ncia}}]{Patra:2025xtd}%
  \BibitemOpen
  \bibfield  {author} {\bibinfo {author} {\bibfnamefont {N.~K.}\ \bibnamefont {Patra}}, \bibinfo {author} {\bibfnamefont {T.}~\bibnamefont {Malik}}, \bibinfo {author} {\bibfnamefont {H.}~\bibnamefont {Pais}}, \bibinfo {author} {\bibfnamefont {K.}~\bibnamefont {Zhou}}, \bibinfo {author} {\bibfnamefont {B.~K.}\ \bibnamefont {Agrawal}}, \ and\ \bibinfo {author} {\bibfnamefont {C.}~\bibnamefont {Provid{\^e}ncia}},\ }\href {\doibase 10.1016/j.physletb.2025.139470} {\bibfield  {journal} {\bibinfo  {journal} {Phys. Lett. B}\ }\textbf {\bibinfo {volume} {865}},\ \bibinfo {pages} {139470} (\bibinfo {year} {2025})},\ \Eprint {http://arxiv.org/abs/2502.20226} {arXiv:2502.20226 [nucl-th]} \BibitemShut {NoStop}%
\bibitem [{\citenamefont {Wang}\ \emph {et~al.}(2025)\citenamefont {Wang}, \citenamefont {Deng}, \citenamefont {Xie}, \citenamefont {Li},\ and\ \citenamefont {Ma}}]{Wang:2024ahj}%
  \BibitemOpen
  \bibfield  {author} {\bibinfo {author} {\bibfnamefont {J.~M.}\ \bibnamefont {Wang}}, \bibinfo {author} {\bibfnamefont {X.~G.}\ \bibnamefont {Deng}}, \bibinfo {author} {\bibfnamefont {W.~J.}\ \bibnamefont {Xie}}, \bibinfo {author} {\bibfnamefont {B.~A.}\ \bibnamefont {Li}}, \ and\ \bibinfo {author} {\bibfnamefont {Y.~G.}\ \bibnamefont {Ma}},\ }\href {\doibase 10.1088/1674-1137/adf4a1} {\bibfield  {journal} {\bibinfo  {journal} {Chin. Phys. C}\ }\textbf {\bibinfo {volume} {49}},\ \bibinfo {pages} {124105} (\bibinfo {year} {2025})},\ \Eprint {http://arxiv.org/abs/2406.07051} {arXiv:2406.07051 [nucl-th]} \BibitemShut {NoStop}%
\bibitem [{\citenamefont {Pu}\ \emph {et~al.}(2023)\citenamefont {Pu}, \citenamefont {Li}, \citenamefont {Lu},\ and\ \citenamefont {Pang}}]{Pu:2023jae}%
  \BibitemOpen
  \bibfield  {author} {\bibinfo {author} {\bibfnamefont {K.~F.}\ \bibnamefont {Pu}}, \bibinfo {author} {\bibfnamefont {H.~L.}\ \bibnamefont {Li}}, \bibinfo {author} {\bibfnamefont {H.~L.}\ \bibnamefont {Lu}}, \ and\ \bibinfo {author} {\bibfnamefont {L.~G.}\ \bibnamefont {Pang}},\ }\href {\doibase 10.1088/1674-1137/acc518} {\bibfield  {journal} {\bibinfo  {journal} {Chin. Phys. C}\ }\textbf {\bibinfo {volume} {47}},\ \bibinfo {pages} {054104} (\bibinfo {year} {2023})},\ \Eprint {http://arxiv.org/abs/2303.03934} {arXiv:2303.03934 [nucl-th]} \BibitemShut {NoStop}%
\bibitem [{\citenamefont {Gao}\ \emph {et~al.}(2024)\citenamefont {Gao}, \citenamefont {Liu}, \citenamefont {Wen}, \citenamefont {Liao}, \citenamefont {Yang}, \citenamefont {Su}, \citenamefont {Wang},\ and\ \citenamefont {Zhu}}]{Gao:2023jjf}%
  \BibitemOpen
  \bibfield  {author} {\bibinfo {author} {\bibfnamefont {Z.~P.}\ \bibnamefont {Gao}}, \bibinfo {author} {\bibfnamefont {S.~Y.}\ \bibnamefont {Liu}}, \bibinfo {author} {\bibfnamefont {P.~W.}\ \bibnamefont {Wen}}, \bibinfo {author} {\bibfnamefont {Z.~H.}\ \bibnamefont {Liao}}, \bibinfo {author} {\bibfnamefont {Y.}~\bibnamefont {Yang}}, \bibinfo {author} {\bibfnamefont {J.}~\bibnamefont {Su}}, \bibinfo {author} {\bibfnamefont {Y.~J.}\ \bibnamefont {Wang}}, \ and\ \bibinfo {author} {\bibfnamefont {L.}~\bibnamefont {Zhu}},\ }\href {\doibase 10.1103/PhysRevC.109.024601} {\bibfield  {journal} {\bibinfo  {journal} {Phys. Rev. C}\ }\textbf {\bibinfo {volume} {109}},\ \bibinfo {pages} {024601} (\bibinfo {year} {2024})},\ \Eprint {http://arxiv.org/abs/2306.11236} {arXiv:2306.11236 [nucl-th]} \BibitemShut {NoStop}%
\bibitem [{\citenamefont {Wu}\ \emph {et~al.}(2025)\citenamefont {Wu}, \citenamefont {Liu}, \citenamefont {Li},\ and\ \citenamefont {Wang}}]{Wu:2025txu}%
  \BibitemOpen
  \bibfield  {author} {\bibinfo {author} {\bibfnamefont {Y.~Q.}\ \bibnamefont {Wu}}, \bibinfo {author} {\bibfnamefont {X.}~\bibnamefont {Liu}}, \bibinfo {author} {\bibfnamefont {H.~L.}\ \bibnamefont {Li}}, \ and\ \bibinfo {author} {\bibfnamefont {F.~Q.}\ \bibnamefont {Wang}},\ }\href@noop {} {\  (\bibinfo {year} {2025})},\ \Eprint {http://arxiv.org/abs/2505.24194} {arXiv:2505.24194 [physics.comp-ph]} \BibitemShut {NoStop}%
\bibitem [{\citenamefont {DONG}\ and\ \citenamefont {GENG}(2023)}]{dong2023machine}%
  \BibitemOpen
  \bibfield  {author} {\bibinfo {author} {\bibfnamefont {X.}~\bibnamefont {DONG}}\ and\ \bibinfo {author} {\bibfnamefont {L.}~\bibnamefont {GENG}},\ }\href {\doibase 10.7538/yzk.2022.youxian.0859} {\bibfield  {journal} {\bibinfo  {journal} {Atomic Energy Science and Technology}\ }\textbf {\bibinfo {volume} {57}},\ \bibinfo {pages} {679} (\bibinfo {year} {2023})}\BibitemShut {NoStop}%
\bibitem [{\citenamefont {Li}\ \emph {et~al.}(2025)\citenamefont {Li}, \citenamefont {Wang}, \citenamefont {Li},\ and\ \citenamefont {Lv}}]{li2025machine}%
  \BibitemOpen
  \bibfield  {author} {\bibinfo {author} {\bibfnamefont {Z.}~\bibnamefont {Li}}, \bibinfo {author} {\bibfnamefont {Y.}~\bibnamefont {Wang}}, \bibinfo {author} {\bibfnamefont {Q.}~\bibnamefont {Li}}, \ and\ \bibinfo {author} {\bibfnamefont {B.-f.}\ \bibnamefont {Lv}},\ }\href {\doibase 10.1103/vj25-zwd3} {\bibfield  {journal} {\bibinfo  {journal} {Phys. Rev. C}\ }\textbf {\bibinfo {volume} {112}},\ \bibinfo {pages} {014312} (\bibinfo {year} {2025})}\BibitemShut {NoStop}%
\bibitem [{\citenamefont {Guo}\ \emph {et~al.}(2026)\citenamefont {Guo}, \citenamefont {Chi},\ and\ \citenamefont {Liu}}]{Guo:2026bfz}%
  \BibitemOpen
  \bibfield  {author} {\bibinfo {author} {\bibfnamefont {X.}~\bibnamefont {Guo}}, \bibinfo {author} {\bibfnamefont {Z.}~\bibnamefont {Chi}}, \ and\ \bibinfo {author} {\bibfnamefont {J.}~\bibnamefont {Liu}},\ }\href {\doibase 10.1142/S0218301326500072} {\bibfield  {journal} {\bibinfo  {journal} {Int. J. Mod. Phys. E}\ }\textbf {\bibinfo {volume} {35}},\ \bibinfo {pages} {2650007} (\bibinfo {year} {2026})}\BibitemShut {NoStop}%
\bibitem [{\citenamefont {Li}\ \emph {et~al.}(2024{\natexlab{a}})\citenamefont {Li}, \citenamefont {Zhang},\ and\ \citenamefont {Fang}}]{Li:2024tth}%
  \BibitemOpen
  \bibfield  {author} {\bibinfo {author} {\bibfnamefont {W.}~\bibnamefont {Li}}, \bibinfo {author} {\bibfnamefont {X.}~\bibnamefont {Zhang}}, \ and\ \bibinfo {author} {\bibfnamefont {J.}~\bibnamefont {Fang}},\ }\href {\doibase 10.1088/1402-4896/ad3170} {\bibfield  {journal} {\bibinfo  {journal} {Phys. Scripta}\ }\textbf {\bibinfo {volume} {99}},\ \bibinfo {pages} {045308} (\bibinfo {year} {2024}{\natexlab{a}})}\BibitemShut {NoStop}%
\bibitem [{\citenamefont {Li}\ \emph {et~al.}(2026)\citenamefont {Li}, \citenamefont {Lv}, \citenamefont {Wang},\ and\ \citenamefont {Petrache}}]{li2026study}%
  \BibitemOpen
  \bibfield  {author} {\bibinfo {author} {\bibfnamefont {Z.~L.}\ \bibnamefont {Li}}, \bibinfo {author} {\bibfnamefont {B.~F.}\ \bibnamefont {Lv}}, \bibinfo {author} {\bibfnamefont {Y.~J.}\ \bibnamefont {Wang}}, \ and\ \bibinfo {author} {\bibfnamefont {C.}~\bibnamefont {Petrache}},\ }\href {\doibase 10.1088/1674-1137/adfe54} {\bibfield  {journal} {\bibinfo  {journal} {Chin. Phys. C}\ }\textbf {\bibinfo {volume} {50}},\ \bibinfo {pages} {014107} (\bibinfo {year} {2026})}\BibitemShut {NoStop}%
\bibitem [{\citenamefont {Du}\ \emph {et~al.}(2023)\citenamefont {Du}, \citenamefont {Guo}, \citenamefont {Wu},\ and\ \citenamefont {Zhang}}]{Du:2023hcq}%
  \BibitemOpen
  \bibfield  {author} {\bibinfo {author} {\bibfnamefont {X.-K.}\ \bibnamefont {Du}}, \bibinfo {author} {\bibfnamefont {P.}~\bibnamefont {Guo}}, \bibinfo {author} {\bibfnamefont {X.-H.}\ \bibnamefont {Wu}}, \ and\ \bibinfo {author} {\bibfnamefont {S.-Q.}\ \bibnamefont {Zhang}},\ }\href {\doibase 10.1088/1674-1137/acc791} {\bibfield  {journal} {\bibinfo  {journal} {Chin. Phys. C}\ }\textbf {\bibinfo {volume} {47}},\ \bibinfo {pages} {074108} (\bibinfo {year} {2023})}\BibitemShut {NoStop}%
\bibitem [{\citenamefont {Huang}\ \emph {et~al.}(2025)\citenamefont {Huang}, \citenamefont {Wendt}, \citenamefont {Schunck},\ and\ \citenamefont {Holmbeck}}]{Huang:2025ubz}%
  \BibitemOpen
  \bibfield  {author} {\bibinfo {author} {\bibfnamefont {M.}~\bibnamefont {Huang}}, \bibinfo {author} {\bibfnamefont {K.~A.}\ \bibnamefont {Wendt}}, \bibinfo {author} {\bibfnamefont {N.~F.}\ \bibnamefont {Schunck}}, \ and\ \bibinfo {author} {\bibfnamefont {E.~M.}\ \bibnamefont {Holmbeck}},\ }\href {\doibase 10.1103/tdk4-c4tp} {\bibfield  {journal} {\bibinfo  {journal} {Phys. Rev. C}\ }\textbf {\bibinfo {volume} {112}},\ \bibinfo {pages} {034317} (\bibinfo {year} {2025})},\ \Eprint {http://arxiv.org/abs/2504.09013} {arXiv:2504.09013 [nucl-th]} \BibitemShut {NoStop}%
\bibitem [{\citenamefont {Bentley}\ \emph {et~al.}(2025)\citenamefont {Bentley}, \citenamefont {Tedder}, \citenamefont {Gebran},\ and\ \citenamefont {Paul}}]{Bentley:2024bnm}%
  \BibitemOpen
  \bibfield  {author} {\bibinfo {author} {\bibfnamefont {I.}~\bibnamefont {Bentley}}, \bibinfo {author} {\bibfnamefont {J.}~\bibnamefont {Tedder}}, \bibinfo {author} {\bibfnamefont {M.}~\bibnamefont {Gebran}}, \ and\ \bibinfo {author} {\bibfnamefont {A.}~\bibnamefont {Paul}},\ }\href {\doibase 10.1103/PhysRevC.111.034305} {\bibfield  {journal} {\bibinfo  {journal} {Phys. Rev. C}\ }\textbf {\bibinfo {volume} {111}},\ \bibinfo {pages} {034305} (\bibinfo {year} {2025})},\ \Eprint {http://arxiv.org/abs/2412.09504} {arXiv:2412.09504 [nucl-th]} \BibitemShut {NoStop}%
\bibitem [{\citenamefont {Yuan}\ \emph {et~al.}(2024)\citenamefont {Yuan}, \citenamefont {Bai}, \citenamefont {Wang},\ and\ \citenamefont {Ren}}]{Yuan:2024ivv}%
  \BibitemOpen
  \bibfield  {author} {\bibinfo {author} {\bibfnamefont {Z.-Y.}\ \bibnamefont {Yuan}}, \bibinfo {author} {\bibfnamefont {D.}~\bibnamefont {Bai}}, \bibinfo {author} {\bibfnamefont {Z.}~\bibnamefont {Wang}}, \ and\ \bibinfo {author} {\bibfnamefont {Z.-Z.}\ \bibnamefont {Ren}},\ }\href {\doibase 10.1007/s41365-024-01463-9} {\bibfield  {journal} {\bibinfo  {journal} {Nucl. Sci. Tech.}\ }\textbf {\bibinfo {volume} {35}},\ \bibinfo {pages} {105} (\bibinfo {year} {2024})}\BibitemShut {NoStop}%
\bibitem [{\citenamefont {Jyothish}\ \emph {et~al.}(2025)\citenamefont {Jyothish}, \citenamefont {Manangode},\ and\ \citenamefont {Rhine~Kumar}}]{Jyothish:2025ddz}%
  \BibitemOpen
  \bibfield  {author} {\bibinfo {author} {\bibfnamefont {K.}~\bibnamefont {Jyothish}}, \bibinfo {author} {\bibfnamefont {G.}~\bibnamefont {Manangode}}, \ and\ \bibinfo {author} {\bibfnamefont {A.~K.}\ \bibnamefont {Rhine~Kumar}},\ }\href {\doibase 10.1103/67c1-2dvd} {\bibfield  {journal} {\bibinfo  {journal} {Phys. Rev. C}\ }\textbf {\bibinfo {volume} {112}},\ \bibinfo {pages} {064309} (\bibinfo {year} {2025})}\BibitemShut {NoStop}%
\bibitem [{\citenamefont {Cai}\ and\ \citenamefont {Yuan}(2023)}]{Cai:2023mak}%
  \BibitemOpen
  \bibfield  {author} {\bibinfo {author} {\bibfnamefont {B.-S.}\ \bibnamefont {Cai}}\ and\ \bibinfo {author} {\bibfnamefont {C.-X.}\ \bibnamefont {Yuan}},\ }\href {\doibase 10.1007/s41365-023-01354-5} {\bibfield  {journal} {\bibinfo  {journal} {Nucl. Sci. Tech.}\ }\textbf {\bibinfo {volume} {34}},\ \bibinfo {pages} {204} (\bibinfo {year} {2023})},\ \Eprint {http://arxiv.org/abs/2305.05209} {arXiv:2305.05209 [nucl-th]} \BibitemShut {NoStop}%
\bibitem [{\citenamefont {Ma}\ \emph {et~al.}(2023)\citenamefont {Ma}, \citenamefont {Zhao}, \citenamefont {Wang},\ and\ \citenamefont {Zhang}}]{Ma:2023ofi}%
  \BibitemOpen
  \bibfield  {author} {\bibinfo {author} {\bibfnamefont {N.~N.}\ \bibnamefont {Ma}}, \bibinfo {author} {\bibfnamefont {T.~L.}\ \bibnamefont {Zhao}}, \bibinfo {author} {\bibfnamefont {W.~X.}\ \bibnamefont {Wang}}, \ and\ \bibinfo {author} {\bibfnamefont {H.~F.}\ \bibnamefont {Zhang}},\ }\href {\doibase 10.1103/PhysRevC.107.014310} {\bibfield  {journal} {\bibinfo  {journal} {Phys. Rev. C}\ }\textbf {\bibinfo {volume} {107}},\ \bibinfo {pages} {014310} (\bibinfo {year} {2023})}\BibitemShut {NoStop}%
\bibitem [{\citenamefont {Tang}\ and\ \citenamefont {Zhang}(2024)}]{Tang:2024xxj}%
  \BibitemOpen
  \bibfield  {author} {\bibinfo {author} {\bibfnamefont {L.}~\bibnamefont {Tang}}\ and\ \bibinfo {author} {\bibfnamefont {Z.~H.}\ \bibnamefont {Zhang}},\ }\href {\doibase 10.1007/s41365-024-01379-4} {\bibfield  {journal} {\bibinfo  {journal} {Nucl. Sci. Tech.}\ }\textbf {\bibinfo {volume} {35}},\ \bibinfo {pages} {19} (\bibinfo {year} {2024})},\ \Eprint {http://arxiv.org/abs/2404.12609} {arXiv:2404.12609 [nucl-th]} \BibitemShut {NoStop}%
\bibitem [{\citenamefont {Shree}\ and\ \citenamefont {Balasubramaniam}(2025)}]{Shree:2025hyu}%
  \BibitemOpen
  \bibfield  {author} {\bibinfo {author} {\bibfnamefont {S.~M.}\ \bibnamefont {Shree}}\ and\ \bibinfo {author} {\bibfnamefont {M.}~\bibnamefont {Balasubramaniam}},\ }\href {\doibase 10.1140/epja/s10050-025-01494-9} {\bibfield  {journal} {\bibinfo  {journal} {Eur. Phys. J. A}\ }\textbf {\bibinfo {volume} {61}},\ \bibinfo {pages} {32} (\bibinfo {year} {2025})}\BibitemShut {NoStop}%
\bibitem [{\citenamefont {Niu}\ \emph {et~al.}(2019)\citenamefont {Niu}, \citenamefont {Liang}, \citenamefont {Sun}, \citenamefont {Long},\ and\ \citenamefont {Niu}}]{Niu:2018trk}%
  \BibitemOpen
  \bibfield  {author} {\bibinfo {author} {\bibfnamefont {Z.~M.}\ \bibnamefont {Niu}}, \bibinfo {author} {\bibfnamefont {H.~Z.}\ \bibnamefont {Liang}}, \bibinfo {author} {\bibfnamefont {B.~H.}\ \bibnamefont {Sun}}, \bibinfo {author} {\bibfnamefont {W.~H.}\ \bibnamefont {Long}}, \ and\ \bibinfo {author} {\bibfnamefont {Y.~F.}\ \bibnamefont {Niu}},\ }\href {\doibase 10.1103/PhysRevC.99.064307} {\bibfield  {journal} {\bibinfo  {journal} {Phys. Rev. C}\ }\textbf {\bibinfo {volume} {99}},\ \bibinfo {pages} {064307} (\bibinfo {year} {2019})},\ \Eprint {http://arxiv.org/abs/1810.03156} {arXiv:1810.03156 [nucl-th]} \BibitemShut {NoStop}%
\bibitem [{\citenamefont {Li}\ \emph {et~al.}(2024{\natexlab{b}})\citenamefont {Li}, \citenamefont {Gao}, \citenamefont {Liu}, \citenamefont {Wang}, \citenamefont {Zhu},\ and\ \citenamefont {Li}}]{Li:2023ukd}%
  \BibitemOpen
  \bibfield  {author} {\bibinfo {author} {\bibfnamefont {Z.}~\bibnamefont {Li}}, \bibinfo {author} {\bibfnamefont {Z.}~\bibnamefont {Gao}}, \bibinfo {author} {\bibfnamefont {L.}~\bibnamefont {Liu}}, \bibinfo {author} {\bibfnamefont {Y.}~\bibnamefont {Wang}}, \bibinfo {author} {\bibfnamefont {L.}~\bibnamefont {Zhu}}, \ and\ \bibinfo {author} {\bibfnamefont {Q.}~\bibnamefont {Li}},\ }\href {\doibase 10.1103/PhysRevC.109.024604} {\bibfield  {journal} {\bibinfo  {journal} {Phys. Rev. C}\ }\textbf {\bibinfo {volume} {109}},\ \bibinfo {pages} {024604} (\bibinfo {year} {2024}{\natexlab{b}})},\ \Eprint {http://arxiv.org/abs/2310.04700} {arXiv:2310.04700 [nucl-th]} \BibitemShut {NoStop}%
\bibitem [{\citenamefont {Choi}\ \emph {et~al.}(2025)\citenamefont {Choi}, \citenamefont {Mitra}, \citenamefont {Brodksy}, \citenamefont {Glatt}, \citenamefont {Holmbeck}, \citenamefont {Liu}, \citenamefont {Schunck}, \citenamefont {Sieverding},\ and\ \citenamefont {Wendt}}]{Choi:2024nkr}%
  \BibitemOpen
  \bibfield  {author} {\bibinfo {author} {\bibfnamefont {H.}~\bibnamefont {Choi}}, \bibinfo {author} {\bibfnamefont {S.}~\bibnamefont {Mitra}}, \bibinfo {author} {\bibfnamefont {J.}~\bibnamefont {Brodksy}}, \bibinfo {author} {\bibfnamefont {R.}~\bibnamefont {Glatt}}, \bibinfo {author} {\bibfnamefont {E.}~\bibnamefont {Holmbeck}}, \bibinfo {author} {\bibfnamefont {S.}~\bibnamefont {Liu}}, \bibinfo {author} {\bibfnamefont {N.}~\bibnamefont {Schunck}}, \bibinfo {author} {\bibfnamefont {A.}~\bibnamefont {Sieverding}}, \ and\ \bibinfo {author} {\bibfnamefont {K.}~\bibnamefont {Wendt}},\ }\href {\doibase 10.1103/4jd4-bnyh} {\bibfield  {journal} {\bibinfo  {journal} {Phys. Rev. C}\ }\textbf {\bibinfo {volume} {112}},\ \bibinfo {pages} {044601} (\bibinfo {year} {2025})},\ \Eprint {http://arxiv.org/abs/2404.02332} {arXiv:2404.02332 [nucl-th]} \BibitemShut {NoStop}%
\bibitem [{\citenamefont {Bass}\ \emph {et~al.}(1994)\citenamefont {Bass}, \citenamefont {Bischoff}, \citenamefont {Hartnack}, \citenamefont {Maruhn}, \citenamefont {Reinhardt}, \citenamefont {Stoecker},\ and\ \citenamefont {Greiner}}]{Bass:1993vx}%
  \BibitemOpen
  \bibfield  {author} {\bibinfo {author} {\bibfnamefont {S.~A.}\ \bibnamefont {Bass}}, \bibinfo {author} {\bibfnamefont {A.}~\bibnamefont {Bischoff}}, \bibinfo {author} {\bibfnamefont {C.}~\bibnamefont {Hartnack}}, \bibinfo {author} {\bibfnamefont {J.~A.}\ \bibnamefont {Maruhn}}, \bibinfo {author} {\bibfnamefont {J.}~\bibnamefont {Reinhardt}}, \bibinfo {author} {\bibfnamefont {H.}~\bibnamefont {Stoecker}}, \ and\ \bibinfo {author} {\bibfnamefont {W.}~\bibnamefont {Greiner}},\ }\href {\doibase 10.1088/0954-3899/20/1/004} {\bibfield  {journal} {\bibinfo  {journal} {J. Phys. G}\ }\textbf {\bibinfo {volume} {20}},\ \bibinfo {pages} {L21} (\bibinfo {year} {1994})}\BibitemShut {NoStop}%
\bibitem [{\citenamefont {Bass}\ \emph {et~al.}(1996)\citenamefont {Bass}, \citenamefont {Bischoff}, \citenamefont {Maruhn}, \citenamefont {Stoecker},\ and\ \citenamefont {Greiner}}]{Bass:1996ez}%
  \BibitemOpen
  \bibfield  {author} {\bibinfo {author} {\bibfnamefont {S.~A.}\ \bibnamefont {Bass}}, \bibinfo {author} {\bibfnamefont {A.}~\bibnamefont {Bischoff}}, \bibinfo {author} {\bibfnamefont {J.~A.}\ \bibnamefont {Maruhn}}, \bibinfo {author} {\bibfnamefont {H.}~\bibnamefont {Stoecker}}, \ and\ \bibinfo {author} {\bibfnamefont {W.}~\bibnamefont {Greiner}},\ }\href {\doibase 10.1103/PhysRevC.53.2358} {\bibfield  {journal} {\bibinfo  {journal} {Phys. Rev. C}\ }\textbf {\bibinfo {volume} {53}},\ \bibinfo {pages} {2358} (\bibinfo {year} {1996})},\ \Eprint {http://arxiv.org/abs/nucl-th/9601024} {arXiv:nucl-th/9601024} \BibitemShut {NoStop}%
\bibitem [{\citenamefont {De~Sanctis}\ \emph {et~al.}(2009)\citenamefont {De~Sanctis}, \citenamefont {Masotti}, \citenamefont {Bruno}, \citenamefont {D'Agostino}, \citenamefont {Geraci}, \citenamefont {Vannini},\ and\ \citenamefont {Bonasera}}]{DeSanctis:2009zzb}%
  \BibitemOpen
  \bibfield  {author} {\bibinfo {author} {\bibfnamefont {J.}~\bibnamefont {De~Sanctis}}, \bibinfo {author} {\bibfnamefont {M.}~\bibnamefont {Masotti}}, \bibinfo {author} {\bibfnamefont {M.}~\bibnamefont {Bruno}}, \bibinfo {author} {\bibfnamefont {M.}~\bibnamefont {D'Agostino}}, \bibinfo {author} {\bibfnamefont {E.}~\bibnamefont {Geraci}}, \bibinfo {author} {\bibfnamefont {G.}~\bibnamefont {Vannini}}, \ and\ \bibinfo {author} {\bibfnamefont {A.}~\bibnamefont {Bonasera}},\ }\href {\doibase 10.1088/0954-3899/36/1/015101} {\bibfield  {journal} {\bibinfo  {journal} {J. Phys. G}\ }\textbf {\bibinfo {volume} {36}},\ \bibinfo {pages} {015101} (\bibinfo {year} {2009})}\BibitemShut {NoStop}%
\bibitem [{\citenamefont {David}\ \emph {et~al.}(1995)\citenamefont {David}, \citenamefont {Freslier},\ and\ \citenamefont {Aichelin}}]{David:1994qc}%
  \BibitemOpen
  \bibfield  {author} {\bibinfo {author} {\bibfnamefont {C.}~\bibnamefont {David}}, \bibinfo {author} {\bibfnamefont {M.}~\bibnamefont {Freslier}}, \ and\ \bibinfo {author} {\bibfnamefont {J.}~\bibnamefont {Aichelin}},\ }\href {\doibase 10.1103/PhysRevC.51.1453} {\bibfield  {journal} {\bibinfo  {journal} {Phys. Rev. C}\ }\textbf {\bibinfo {volume} {51}},\ \bibinfo {pages} {1453} (\bibinfo {year} {1995})}\BibitemShut {NoStop}%
\bibitem [{\citenamefont {Galaktionov}\ \emph {et~al.}(2023{\natexlab{a}})\citenamefont {Galaktionov}, \citenamefont {Roudnev},\ and\ \citenamefont {Valiev}}]{Galaktionov:2023dui}%
  \BibitemOpen
  \bibfield  {author} {\bibinfo {author} {\bibfnamefont {K.~A.}\ \bibnamefont {Galaktionov}}, \bibinfo {author} {\bibfnamefont {V.~A.}\ \bibnamefont {Roudnev}}, \ and\ \bibinfo {author} {\bibfnamefont {F.~F.}\ \bibnamefont {Valiev}},\ }\href {\doibase 10.1134/S1063778823060248} {\bibfield  {journal} {\bibinfo  {journal} {Phys. Atom. Nucl.}\ }\textbf {\bibinfo {volume} {86}},\ \bibinfo {pages} {1426} (\bibinfo {year} {2023}{\natexlab{a}})}\BibitemShut {NoStop}%
\bibitem [{\citenamefont {Galaktionov}\ \emph {et~al.}(2023{\natexlab{b}})\citenamefont {Galaktionov}, \citenamefont {Rudnev},\ and\ \citenamefont {Valiev}}]{Galaktionov:2023ilx}%
  \BibitemOpen
  \bibfield  {author} {\bibinfo {author} {\bibfnamefont {K.}~\bibnamefont {Galaktionov}}, \bibinfo {author} {\bibfnamefont {V.}~\bibnamefont {Rudnev}}, \ and\ \bibinfo {author} {\bibfnamefont {F.}~\bibnamefont {Valiev}},\ }\href {\doibase 10.1134/S1063779623030152} {\bibfield  {journal} {\bibinfo  {journal} {Phys. Part. Nucl.}\ }\textbf {\bibinfo {volume} {54}},\ \bibinfo {pages} {446} (\bibinfo {year} {2023}{\natexlab{b}})}\BibitemShut {NoStop}%
\bibitem [{\citenamefont {Zhang}\ \emph {et~al.}(2022)\citenamefont {Zhang} \emph {et~al.}}]{Zhang:2021zxd}%
  \BibitemOpen
  \bibfield  {author} {\bibinfo {author} {\bibfnamefont {X.}~\bibnamefont {Zhang}} \emph {et~al.},\ }\href {\doibase 10.1103/PhysRevC.105.034611} {\bibfield  {journal} {\bibinfo  {journal} {Phys. Rev. C}\ }\textbf {\bibinfo {volume} {105}},\ \bibinfo {pages} {034611} (\bibinfo {year} {2022})},\ \Eprint {http://arxiv.org/abs/2111.06597} {arXiv:2111.06597 [nucl-th]} \BibitemShut {NoStop}%
\bibitem [{\citenamefont {Xiang}\ \emph {et~al.}(2022)\citenamefont {Xiang}, \citenamefont {Zhao},\ and\ \citenamefont {Huang}}]{Xiang:2021ssj}%
  \BibitemOpen
  \bibfield  {author} {\bibinfo {author} {\bibfnamefont {P.}~\bibnamefont {Xiang}}, \bibinfo {author} {\bibfnamefont {Y.-S.}\ \bibnamefont {Zhao}}, \ and\ \bibinfo {author} {\bibfnamefont {X.-G.}\ \bibnamefont {Huang}},\ }\href {\doibase 10.1088/1674-1137/ac6490} {\bibfield  {journal} {\bibinfo  {journal} {Chin. Phys. C}\ }\textbf {\bibinfo {volume} {46}},\ \bibinfo {pages} {074110} (\bibinfo {year} {2022})},\ \Eprint {http://arxiv.org/abs/2112.03824} {arXiv:2112.03824 [hep-ph]} \BibitemShut {NoStop}%
\bibitem [{\citenamefont {Mallick}\ \emph {et~al.}(2021)\citenamefont {Mallick}, \citenamefont {Tripathy}, \citenamefont {Mishra}, \citenamefont {Deb},\ and\ \citenamefont {Sahoo}}]{Mallick:2021wop}%
  \BibitemOpen
  \bibfield  {author} {\bibinfo {author} {\bibfnamefont {N.}~\bibnamefont {Mallick}}, \bibinfo {author} {\bibfnamefont {S.}~\bibnamefont {Tripathy}}, \bibinfo {author} {\bibfnamefont {A.~N.}\ \bibnamefont {Mishra}}, \bibinfo {author} {\bibfnamefont {S.}~\bibnamefont {Deb}}, \ and\ \bibinfo {author} {\bibfnamefont {R.}~\bibnamefont {Sahoo}},\ }\href {\doibase 10.1103/PhysRevD.103.094031} {\bibfield  {journal} {\bibinfo  {journal} {Phys. Rev. D}\ }\textbf {\bibinfo {volume} {103}},\ \bibinfo {pages} {094031} (\bibinfo {year} {2021})},\ \Eprint {http://arxiv.org/abs/2103.01736} {arXiv:2103.01736 [hep-ph]} \BibitemShut {NoStop}%
\bibitem [{\citenamefont {LI}\ \emph {et~al.}(2022)\citenamefont {LI}, \citenamefont {ZHANG}, \citenamefont {CUI},\ and\ \citenamefont {LIANG}}]{LiLI:2022vlp}%
  \BibitemOpen
  \bibfield  {author} {\bibinfo {author} {\bibfnamefont {L.}~\bibnamefont {LI}}, \bibinfo {author} {\bibfnamefont {Y.}~\bibnamefont {ZHANG}}, \bibinfo {author} {\bibfnamefont {Y.}~\bibnamefont {CUI}}, \ and\ \bibinfo {author} {\bibfnamefont {J.}~\bibnamefont {LIANG}},\ }\href {\doibase 10.1360/SSPMA-2021-0303} {\bibfield  {journal} {\bibinfo  {journal} {Sci. Sin. Phys. Mech. Astro.}\ }\textbf {\bibinfo {volume} {52}},\ \bibinfo {pages} {252014} (\bibinfo {year} {2022})}\BibitemShut {NoStop}%
\bibitem [{\citenamefont {Li}\ \emph {et~al.}(2021)\citenamefont {Li}, \citenamefont {Wang}, \citenamefont {Gao}, \citenamefont {Li}, \citenamefont {L{\"u}}, \citenamefont {Lv}, \citenamefont {Li}, \citenamefont {Tsang},\ and\ \citenamefont {Tsang}}]{Li:2021plq}%
  \BibitemOpen
  \bibfield  {author} {\bibinfo {author} {\bibfnamefont {F.}~\bibnamefont {Li}}, \bibinfo {author} {\bibfnamefont {Y.}~\bibnamefont {Wang}}, \bibinfo {author} {\bibfnamefont {Z.}~\bibnamefont {Gao}}, \bibinfo {author} {\bibfnamefont {P.}~\bibnamefont {Li}}, \bibinfo {author} {\bibfnamefont {H.}~\bibnamefont {L{\"u}}}, \bibinfo {author} {\bibfnamefont {H.}~\bibnamefont {Lv}}, \bibinfo {author} {\bibfnamefont {Q.}~\bibnamefont {Li}}, \bibinfo {author} {\bibfnamefont {C.~Y.}\ \bibnamefont {Tsang}}, \ and\ \bibinfo {author} {\bibfnamefont {M.~B.}\ \bibnamefont {Tsang}},\ }\href {\doibase 10.1103/PhysRevC.104.034608} {\bibfield  {journal} {\bibinfo  {journal} {Phys. Rev. C}\ }\textbf {\bibinfo {volume} {104}},\ \bibinfo {pages} {034608} (\bibinfo {year} {2021})},\ \Eprint {http://arxiv.org/abs/2105.08912} {arXiv:2105.08912 [nucl-th]} \BibitemShut {NoStop}%
\bibitem [{\citenamefont {Li}\ \emph {et~al.}(2020)\citenamefont {Li}, \citenamefont {Wang}, \citenamefont {L{\"u}}, \citenamefont {Li}, \citenamefont {Li},\ and\ \citenamefont {Liu}}]{Li:2020qqn}%
  \BibitemOpen
  \bibfield  {author} {\bibinfo {author} {\bibfnamefont {F.}~\bibnamefont {Li}}, \bibinfo {author} {\bibfnamefont {Y.}~\bibnamefont {Wang}}, \bibinfo {author} {\bibfnamefont {H.}~\bibnamefont {L{\"u}}}, \bibinfo {author} {\bibfnamefont {P.}~\bibnamefont {Li}}, \bibinfo {author} {\bibfnamefont {Q.}~\bibnamefont {Li}}, \ and\ \bibinfo {author} {\bibfnamefont {F.}~\bibnamefont {Liu}},\ }\href {\doibase 10.1088/1361-6471/abb1f9} {\bibfield  {journal} {\bibinfo  {journal} {J. Phys. G}\ }\textbf {\bibinfo {volume} {47}},\ \bibinfo {pages} {115104} (\bibinfo {year} {2020})},\ \Eprint {http://arxiv.org/abs/2008.11540} {arXiv:2008.11540 [nucl-th]} \BibitemShut {NoStop}%
\bibitem [{\citenamefont {Li}\ \emph {et~al.}(2011)\citenamefont {Li}, \citenamefont {Shen}, \citenamefont {Guo}, \citenamefont {Wang}, \citenamefont {Li}, \citenamefont {Lukasik},\ and\ \citenamefont {Trautmann}}]{Li:2011zzp}%
  \BibitemOpen
  \bibfield  {author} {\bibinfo {author} {\bibfnamefont {Q.}~\bibnamefont {Li}}, \bibinfo {author} {\bibfnamefont {C.}~\bibnamefont {Shen}}, \bibinfo {author} {\bibfnamefont {C.}~\bibnamefont {Guo}}, \bibinfo {author} {\bibfnamefont {Y.}~\bibnamefont {Wang}}, \bibinfo {author} {\bibfnamefont {Z.}~\bibnamefont {Li}}, \bibinfo {author} {\bibfnamefont {J.}~\bibnamefont {Lukasik}}, \ and\ \bibinfo {author} {\bibfnamefont {W.}~\bibnamefont {Trautmann}},\ }\href {\doibase 10.1103/PhysRevC.83.044617} {\bibfield  {journal} {\bibinfo  {journal} {Phys. Rev. C}\ }\textbf {\bibinfo {volume} {83}},\ \bibinfo {pages} {044617} (\bibinfo {year} {2011})}\BibitemShut {NoStop}%
\bibitem [{\citenamefont {Wang}\ \emph {et~al.}(2021)\citenamefont {Wang}, \citenamefont {Gao},\ and\ \citenamefont {Li}}]{Wang:2021symmetry}%
  \BibitemOpen
  \bibfield  {author} {\bibinfo {author} {\bibfnamefont {Y.}~\bibnamefont {Wang}}, \bibinfo {author} {\bibfnamefont {Z.}~\bibnamefont {Gao}}, \ and\ \bibinfo {author} {\bibfnamefont {Q.}~\bibnamefont {Li}},\ }\href {\doibase 10.3390/sym13112172} {\bibfield  {journal} {\bibinfo  {journal} {Symmetry}\ }\textbf {\bibinfo {volume} {13}},\ \bibinfo {pages} {2172} (\bibinfo {year} {2021})}\BibitemShut {NoStop}%
\bibitem [{\citenamefont {Gao}\ \emph {et~al.}(2023)\citenamefont {Gao}, \citenamefont {Wang}, \citenamefont {Gao},\ and\ \citenamefont {Li}}]{Gao:2023plb}%
  \BibitemOpen
  \bibfield  {author} {\bibinfo {author} {\bibfnamefont {B.}~\bibnamefont {Gao}}, \bibinfo {author} {\bibfnamefont {Y.}~\bibnamefont {Wang}}, \bibinfo {author} {\bibfnamefont {Z.}~\bibnamefont {Gao}}, \ and\ \bibinfo {author} {\bibfnamefont {Q.}~\bibnamefont {Li}},\ }\href {\doibase 10.1016/j.physletb.2023.137685} {\bibfield  {journal} {\bibinfo  {journal} {Phys. Lett. B}\ }\textbf {\bibinfo {volume} {838}},\ \bibinfo {pages} {137685} (\bibinfo {year} {2023})}\BibitemShut {NoStop}%
\bibitem [{\citenamefont {Lin}\ \emph {et~al.}(2005)\citenamefont {Lin}, \citenamefont {Ko}, \citenamefont {Li}, \citenamefont {Zhang},\ and\ \citenamefont {Pal}}]{Lin:2004en}%
  \BibitemOpen
  \bibfield  {author} {\bibinfo {author} {\bibfnamefont {Z.-W.}\ \bibnamefont {Lin}}, \bibinfo {author} {\bibfnamefont {C.~M.}\ \bibnamefont {Ko}}, \bibinfo {author} {\bibfnamefont {B.-A.}\ \bibnamefont {Li}}, \bibinfo {author} {\bibfnamefont {B.}~\bibnamefont {Zhang}}, \ and\ \bibinfo {author} {\bibfnamefont {S.}~\bibnamefont {Pal}},\ }\href {\doibase 10.1103/PhysRevC.72.064901} {\bibfield  {journal} {\bibinfo  {journal} {Phys. Rev. C}\ }\textbf {\bibinfo {volume} {72}},\ \bibinfo {pages} {064901} (\bibinfo {year} {2005})},\ \Eprint {http://arxiv.org/abs/nucl-th/0411110} {arXiv:nucl-th/0411110} \BibitemShut {NoStop}%
\bibitem [{\citenamefont {Nara}\ \emph {et~al.}(1999)\citenamefont {Nara}, \citenamefont {Otuka}, \citenamefont {Ohnishi}, \citenamefont {Niita},\ and\ \citenamefont {Chiba}}]{Nara:1999dz}%
  \BibitemOpen
  \bibfield  {author} {\bibinfo {author} {\bibfnamefont {Y.}~\bibnamefont {Nara}}, \bibinfo {author} {\bibfnamefont {N.}~\bibnamefont {Otuka}}, \bibinfo {author} {\bibfnamefont {A.}~\bibnamefont {Ohnishi}}, \bibinfo {author} {\bibfnamefont {K.}~\bibnamefont {Niita}}, \ and\ \bibinfo {author} {\bibfnamefont {S.}~\bibnamefont {Chiba}},\ }\href {\doibase 10.1103/PhysRevC.61.024901} {\bibfield  {journal} {\bibinfo  {journal} {Phys. Rev. C}\ }\textbf {\bibinfo {volume} {61}},\ \bibinfo {pages} {024901} (\bibinfo {year} {1999})},\ \Eprint {http://arxiv.org/abs/nucl-th/9904059} {arXiv:nucl-th/9904059} \BibitemShut {NoStop}%
\bibitem [{\citenamefont {Isse}\ \emph {et~al.}(2005)\citenamefont {Isse}, \citenamefont {Ohnishi}, \citenamefont {Otuka}, \citenamefont {Sahu},\ and\ \citenamefont {Nara}}]{Isse:2005nk}%
  \BibitemOpen
  \bibfield  {author} {\bibinfo {author} {\bibfnamefont {M.}~\bibnamefont {Isse}}, \bibinfo {author} {\bibfnamefont {A.}~\bibnamefont {Ohnishi}}, \bibinfo {author} {\bibfnamefont {N.}~\bibnamefont {Otuka}}, \bibinfo {author} {\bibfnamefont {P.~K.}\ \bibnamefont {Sahu}}, \ and\ \bibinfo {author} {\bibfnamefont {Y.}~\bibnamefont {Nara}},\ }\href {\doibase 10.1103/PhysRevC.72.064908} {\bibfield  {journal} {\bibinfo  {journal} {Phys. Rev. C}\ }\textbf {\bibinfo {volume} {72}},\ \bibinfo {pages} {064908} (\bibinfo {year} {2005})},\ \Eprint {http://arxiv.org/abs/nucl-th/0502058} {arXiv:nucl-th/0502058} \BibitemShut {NoStop}%
\bibitem [{\citenamefont {Nara}\ and\ \citenamefont {Ohnishi}(2022)}]{Nara:2021fuu}%
  \BibitemOpen
  \bibfield  {author} {\bibinfo {author} {\bibfnamefont {Y.}~\bibnamefont {Nara}}\ and\ \bibinfo {author} {\bibfnamefont {A.}~\bibnamefont {Ohnishi}},\ }\href {\doibase 10.1103/PhysRevC.105.014911} {\bibfield  {journal} {\bibinfo  {journal} {Phys. Rev. C}\ }\textbf {\bibinfo {volume} {105}},\ \bibinfo {pages} {014911} (\bibinfo {year} {2022})},\ \Eprint {http://arxiv.org/abs/2109.07594} {arXiv:2109.07594 [nucl-th]} \BibitemShut {NoStop}%
\bibitem [{\citenamefont {Yue}\ \emph {et~al.}(2022)\citenamefont {Yue}, \citenamefont {Wang}, \citenamefont {Li},\ and\ \citenamefont {Liu}}]{Yue:2022zfu}%
  \BibitemOpen
  \bibfield  {author} {\bibinfo {author} {\bibfnamefont {X.}~\bibnamefont {Yue}}, \bibinfo {author} {\bibfnamefont {Y.}~\bibnamefont {Wang}}, \bibinfo {author} {\bibfnamefont {Q.}~\bibnamefont {Li}}, \ and\ \bibinfo {author} {\bibfnamefont {F.}~\bibnamefont {Liu}},\ }\href {\doibase 10.3390/universe8090491} {\bibfield  {journal} {\bibinfo  {journal} {Universe}\ }\textbf {\bibinfo {volume} {8}},\ \bibinfo {pages} {491} (\bibinfo {year} {2022})},\ \Eprint {http://arxiv.org/abs/2209.10923} {arXiv:2209.10923 [nucl-th]} \BibitemShut {NoStop}%
\bibitem [{\citenamefont {Xu}\ \emph {et~al.}(2014)\citenamefont {Xu}, \citenamefont {Song}, \citenamefont {Ko},\ and\ \citenamefont {Li}}]{Xu:2013sta}%
  \BibitemOpen
  \bibfield  {author} {\bibinfo {author} {\bibfnamefont {J.}~\bibnamefont {Xu}}, \bibinfo {author} {\bibfnamefont {T.}~\bibnamefont {Song}}, \bibinfo {author} {\bibfnamefont {C.~M.}\ \bibnamefont {Ko}}, \ and\ \bibinfo {author} {\bibfnamefont {F.}~\bibnamefont {Li}},\ }\href {\doibase 10.1103/PhysRevLett.112.012301} {\bibfield  {journal} {\bibinfo  {journal} {Phys. Rev. Lett.}\ }\textbf {\bibinfo {volume} {112}},\ \bibinfo {pages} {012301} (\bibinfo {year} {2014})},\ \Eprint {http://arxiv.org/abs/1308.1753} {arXiv:1308.1753 [nucl-th]} \BibitemShut {NoStop}%
\bibitem [{\citenamefont {Nara}\ \emph {et~al.}(2017)\citenamefont {Nara}, \citenamefont {Niemi}, \citenamefont {Steinheimer},\ and\ \citenamefont {St{\"o}cker}}]{Nara:2016hbg}%
  \BibitemOpen
  \bibfield  {author} {\bibinfo {author} {\bibfnamefont {Y.}~\bibnamefont {Nara}}, \bibinfo {author} {\bibfnamefont {H.}~\bibnamefont {Niemi}}, \bibinfo {author} {\bibfnamefont {J.}~\bibnamefont {Steinheimer}}, \ and\ \bibinfo {author} {\bibfnamefont {H.}~\bibnamefont {St{\"o}cker}},\ }\href {\doibase 10.1016/j.physletb.2017.02.020} {\bibfield  {journal} {\bibinfo  {journal} {Phys. Lett. B}\ }\textbf {\bibinfo {volume} {769}},\ \bibinfo {pages} {543} (\bibinfo {year} {2017})},\ \Eprint {http://arxiv.org/abs/1611.08023} {arXiv:1611.08023 [nucl-th]} \BibitemShut {NoStop}%
\bibitem [{\citenamefont {Li}\ \emph {et~al.}(2018)\citenamefont {Li}, \citenamefont {Wang}, \citenamefont {Li}, \citenamefont {Guo},\ and\ \citenamefont {Zhang}}]{Li:2018wpv}%
  \BibitemOpen
  \bibfield  {author} {\bibinfo {author} {\bibfnamefont {P.}~\bibnamefont {Li}}, \bibinfo {author} {\bibfnamefont {Y.}~\bibnamefont {Wang}}, \bibinfo {author} {\bibfnamefont {Q.}~\bibnamefont {Li}}, \bibinfo {author} {\bibfnamefont {C.}~\bibnamefont {Guo}}, \ and\ \bibinfo {author} {\bibfnamefont {H.}~\bibnamefont {Zhang}},\ }\href {\doibase 10.1103/PhysRevC.97.044620} {\bibfield  {journal} {\bibinfo  {journal} {Phys. Rev. C}\ }\textbf {\bibinfo {volume} {97}},\ \bibinfo {pages} {044620} (\bibinfo {year} {2018})},\ \Eprint {http://arxiv.org/abs/1804.04288} {arXiv:1804.04288 [nucl-th]} \BibitemShut {NoStop}%
\bibitem [{\citenamefont {Nan}\ \emph {et~al.}(2025)\citenamefont {Nan}, \citenamefont {Li}, \citenamefont {Zuo},\ and\ \citenamefont {Li}}]{Nan:2024ogc}%
  \BibitemOpen
  \bibfield  {author} {\bibinfo {author} {\bibfnamefont {M.}~\bibnamefont {Nan}}, \bibinfo {author} {\bibfnamefont {P.}~\bibnamefont {Li}}, \bibinfo {author} {\bibfnamefont {W.}~\bibnamefont {Zuo}}, \ and\ \bibinfo {author} {\bibfnamefont {Q.}~\bibnamefont {Li}},\ }\href {\doibase 10.1088/1674-1137/add8fd} {\bibfield  {journal} {\bibinfo  {journal} {Chin. Phys. C}\ }\textbf {\bibinfo {volume} {49}},\ \bibinfo {pages} {094112} (\bibinfo {year} {2025})},\ \Eprint {http://arxiv.org/abs/2412.13497} {arXiv:2412.13497 [nucl-th]} \BibitemShut {NoStop}%
\bibitem [{\citenamefont {Liu}\ \emph {et~al.}(2025)\citenamefont {Liu}, \citenamefont {Yang}, \citenamefont {Wang}, \citenamefont {Li}, \citenamefont {Li}, \citenamefont {Xia},\ and\ \citenamefont {Zhang}}]{Liu:2025pzr}%
  \BibitemOpen
  \bibfield  {author} {\bibinfo {author} {\bibfnamefont {Y.-Y.}\ \bibnamefont {Liu}}, \bibinfo {author} {\bibfnamefont {J.-P.}\ \bibnamefont {Yang}}, \bibinfo {author} {\bibfnamefont {Y.-J.}\ \bibnamefont {Wang}}, \bibinfo {author} {\bibfnamefont {Q.-F.}\ \bibnamefont {Li}}, \bibinfo {author} {\bibfnamefont {Z.-X.}\ \bibnamefont {Li}}, \bibinfo {author} {\bibfnamefont {C.-J.}\ \bibnamefont {Xia}}, \ and\ \bibinfo {author} {\bibfnamefont {Y.-X.}\ \bibnamefont {Zhang}},\ }\href {\doibase 10.1007/s41365-024-01607-x} {\bibfield  {journal} {\bibinfo  {journal} {Nucl. Sci. Tech.}\ }\textbf {\bibinfo {volume} {36}},\ \bibinfo {pages} {45} (\bibinfo {year} {2025})}\BibitemShut {NoStop}%
\bibitem [{\citenamefont {Liu}\ \emph {et~al.}(2021)\citenamefont {Liu}, \citenamefont {Wang}, \citenamefont {Cui}, \citenamefont {Xia}, \citenamefont {Li}, \citenamefont {Chen}, \citenamefont {Li},\ and\ \citenamefont {Zhang}}]{Liu:2020jbg}%
  \BibitemOpen
  \bibfield  {author} {\bibinfo {author} {\bibfnamefont {Y.}~\bibnamefont {Liu}}, \bibinfo {author} {\bibfnamefont {Y.}~\bibnamefont {Wang}}, \bibinfo {author} {\bibfnamefont {Y.}~\bibnamefont {Cui}}, \bibinfo {author} {\bibfnamefont {C.-J.}\ \bibnamefont {Xia}}, \bibinfo {author} {\bibfnamefont {Z.}~\bibnamefont {Li}}, \bibinfo {author} {\bibfnamefont {Y.}~\bibnamefont {Chen}}, \bibinfo {author} {\bibfnamefont {Q.}~\bibnamefont {Li}}, \ and\ \bibinfo {author} {\bibfnamefont {Y.}~\bibnamefont {Zhang}},\ }\href {\doibase 10.1103/PhysRevC.103.014616} {\bibfield  {journal} {\bibinfo  {journal} {Phys. Rev. C}\ }\textbf {\bibinfo {volume} {103}},\ \bibinfo {pages} {014616} (\bibinfo {year} {2021})},\ \Eprint {http://arxiv.org/abs/2006.15861} {arXiv:2006.15861 [nucl-th]} \BibitemShut {NoStop}%
\bibitem [{\citenamefont {Wang}\ \emph {et~al.}(2022)\citenamefont {Wang}, \citenamefont {Gao}, \citenamefont {L\"u},\ and\ \citenamefont {Li}}]{Wang:2022cda}%
  \BibitemOpen
  \bibfield  {author} {\bibinfo {author} {\bibfnamefont {Y.}~\bibnamefont {Wang}}, \bibinfo {author} {\bibfnamefont {Z.}~\bibnamefont {Gao}}, \bibinfo {author} {\bibfnamefont {H.}~\bibnamefont {L\"u}}, \ and\ \bibinfo {author} {\bibfnamefont {Q.}~\bibnamefont {Li}},\ }\href {\doibase 10.1016/j.physletb.2022.137508} {\bibfield  {journal} {\bibinfo  {journal} {Phys. Lett. B}\ }\textbf {\bibinfo {volume} {835}},\ \bibinfo {pages} {137508} (\bibinfo {year} {2022})},\ \Eprint {http://arxiv.org/abs/2208.10681} {arXiv:2208.10681 [nucl-th]} \BibitemShut {NoStop}%
\bibitem [{\citenamefont {Omana~Kuttan}\ \emph {et~al.}(2020)\citenamefont {Omana~Kuttan}, \citenamefont {Steinheimer}, \citenamefont {Zhou}, \citenamefont {Redelbach},\ and\ \citenamefont {Stoecker}}]{OmanaKuttan:2020brq}%
  \BibitemOpen
  \bibfield  {author} {\bibinfo {author} {\bibfnamefont {M.}~\bibnamefont {Omana~Kuttan}}, \bibinfo {author} {\bibfnamefont {J.}~\bibnamefont {Steinheimer}}, \bibinfo {author} {\bibfnamefont {K.}~\bibnamefont {Zhou}}, \bibinfo {author} {\bibfnamefont {A.}~\bibnamefont {Redelbach}}, \ and\ \bibinfo {author} {\bibfnamefont {H.}~\bibnamefont {Stoecker}},\ }\href {\doibase 10.1016/j.physletb.2020.135872} {\bibfield  {journal} {\bibinfo  {journal} {Phys. Lett. B}\ }\textbf {\bibinfo {volume} {811}},\ \bibinfo {pages} {135872} (\bibinfo {year} {2020})},\ \Eprint {http://arxiv.org/abs/2009.01584} {arXiv:2009.01584 [hep-ph]} \BibitemShut {NoStop}%
\bibitem [{\citenamefont {Oleksiyuk}\ \emph {et~al.}(2024)\citenamefont {Oleksiyuk}, \citenamefont {Raine}, \citenamefont {Kr{\"a}mer}, \citenamefont {Voloshynovskiy},\ and\ \citenamefont {Golling}}]{Oleksiyuk:2024hru}%
  \BibitemOpen
  \bibfield  {author} {\bibinfo {author} {\bibfnamefont {I.}~\bibnamefont {Oleksiyuk}}, \bibinfo {author} {\bibfnamefont {J.~A.}\ \bibnamefont {Raine}}, \bibinfo {author} {\bibfnamefont {M.}~\bibnamefont {Kr{\"a}mer}}, \bibinfo {author} {\bibfnamefont {S.}~\bibnamefont {Voloshynovskiy}}, \ and\ \bibinfo {author} {\bibfnamefont {T.}~\bibnamefont {Golling}},\ }\href {\doibase 10.1007/JHEP06(2024)163} {\bibfield  {journal} {\bibinfo  {journal} {JHEP}\ }\textbf {\bibinfo {volume} {06}},\ \bibinfo {pages} {163} (\bibinfo {year} {2024})},\ \Eprint {http://arxiv.org/abs/2402.17714} {arXiv:2402.17714 [hep-ph]} \BibitemShut {NoStop}%
\bibitem [{\citenamefont {Lobato}\ \emph {et~al.}(2022)\citenamefont {Lobato}, \citenamefont {Chimanski},\ and\ \citenamefont {Bertulani}}]{Lobato:2022ukl}%
  \BibitemOpen
  \bibfield  {author} {\bibinfo {author} {\bibfnamefont {R.~V.}\ \bibnamefont {Lobato}}, \bibinfo {author} {\bibfnamefont {E.~V.}\ \bibnamefont {Chimanski}}, \ and\ \bibinfo {author} {\bibfnamefont {C.~A.}\ \bibnamefont {Bertulani}},\ }\href {\doibase 10.1088/1742-6596/2340/1/012014} {\bibfield  {journal} {\bibinfo  {journal} {J. Phys. Conf. Ser.}\ }\textbf {\bibinfo {volume} {2340}},\ \bibinfo {pages} {012014} (\bibinfo {year} {2022})},\ \Eprint {http://arxiv.org/abs/2204.01183} {arXiv:2204.01183 [astro-ph.HE]} \BibitemShut {NoStop}%
\bibitem [{\citenamefont {Zhang}\ \emph {et~al.}(2025)\citenamefont {Zhang}, \citenamefont {Guo}, \citenamefont {Zhu}, \citenamefont {Wang},\ and\ \citenamefont {Ma}}]{Zhang:2025maskpoint}%
  \BibitemOpen
  \bibfield  {author} {\bibinfo {author} {\bibfnamefont {J.-Z.}\ \bibnamefont {Zhang}}, \bibinfo {author} {\bibfnamefont {S.}~\bibnamefont {Guo}}, \bibinfo {author} {\bibfnamefont {L.-L.}\ \bibnamefont {Zhu}}, \bibinfo {author} {\bibfnamefont {L.}~\bibnamefont {Wang}}, \ and\ \bibinfo {author} {\bibfnamefont {G.-L.}\ \bibnamefont {Ma}},\ }\href@noop {} {\  (\bibinfo {year} {2025})},\ \Eprint {http://arxiv.org/abs/2510.06691} {arXiv:2510.06691 [nucl-th]} \BibitemShut {NoStop}%
\end{thebibliography}%

\end{document}